\def\year{2022}\relax
\documentclass[letterpaper]{article} % DO NOT CHANGE THIS
\usepackage{aaai22}  % DO NOT CHANGE THIS
\usepackage{times}  % DO NOT CHANGE THIS
\usepackage{helvet}  % DO NOT CHANGE THIS
\usepackage{courier}  % DO NOT CHANGE THIS
\usepackage[hyphens]{url}  % DO NOT CHANGE THIS
\usepackage{graphicx} % DO NOT CHANGE THIS
\usepackage{natbib}  % DO NOT CHANGE THIS AND DO NOT ADD ANY OPTIONS TO IT
\usepackage{caption} % DO NOT CHANGE THIS AND DO NOT ADD ANY OPTIONS TO IT
\DeclareCaptionStyle{ruled}{labelfont=normalfont,labelsep=colon,strut=off} % DO NOT CHANGE THIS
\usepackage{algorithm}
\usepackage{algpseudocode}
\usepackage{pifont}

\usepackage{newfloat}
\usepackage{listings}
\floatstyle{ruled}
\newfloat{listing}{tb}{lst}{}
\floatname{listing}{Listing}
\usepackage{amsmath,bm}
\usepackage{amsfonts}       % blackboard math symbols
\usepackage{multirow}
\usepackage{booktabs}
\usepackage[table,xcdraw]{xcolor}
\usepackage{xcolor}
\usepackage[hidelinks]{hyperref}

\newtheorem{theorem}{Theorem}[section]
\newtheorem{lemma}[theorem]{Lemma}

\newtheorem{assumption}[theorem]{Assumption}

\DeclareMathOperator{\x}{{\mathbf x}}

\DeclareMathOperator{\R}{{\mathbb R}}

\DeclareMathOperator{\D}{{\mathcal D}}
\DeclareMathOperator{\F}{{\mathcal F}}

\DeclareMathOperator{\cP}{{\mathcal P}}
\DeclareMathOperator{\X}{{\mathcal X}}
\DeclareMathOperator{\Y}{{\mathcal Y}}
\DeclareMathOperator*{\E}{{\mathbb{E}}}

\DeclareMathOperator{\A}{{\mathcal A}}

\title{Harmonizing Non-IID Clients for Federated Medical Image Analysis}
\title{Harmonizing Non-IID Features for Stable Aggregation in Federated Medical Image Analysis}
\title{Harmonizing Local and Global Optimization Drifts under Non-IID Features in Federated Medical Image Analysis}
\title{Harmonizing Client and Server Update Shifts under Non-IID Features in Federated Medical Image Analysis}
\title{HarmoFL: Harmonizing Local and Global Drifts for \\
Federated Heterogeneous Medical Image Analysis}
\title{HarmoFL: Harmonizing Local and Global Drifts in \\ Federated Learning on Heterogeneous Medical Images}
\author {
    Meirui Jiang\textsuperscript{\rm 1 \footnote{Equal contribution. Corresponding: mrjiang@cse.cuhk.edu.hk}},
    Zirui Wang\textsuperscript{\rm 2 \footnotemark[1]},
    Qi Dou\textsuperscript{\rm 1}
}
\affiliations {
    \textsuperscript{\rm 1} Department of Computer Science and Engineering,
    The Chinese University of Hong Kong\\ 
    \textsuperscript{\rm 2} School of Biological Science and Medical Engineering, Beihang University\\
}

\begin{document}

\maketitle

\begin{abstract}
Multiple medical institutions collaboratively training a model using federated learning (FL) has become a promising solution for maximizing the potential of data-driven models, yet the non-independent and identically distributed (non-iid) data in medical images is still an outstanding challenge in real-world practice. The feature heterogeneity caused by diverse scanners or protocols introduces a drift in the learning process, in both local (client) and global (server) optimizations, which harms the convergence as well as model performance. Many previous works have attempted to address the non-iid issue by tackling the drift locally or globally, but how to jointly solve the two essentially coupled drifts is still unclear. In this work, we concentrate on handling both local and global drifts and introduce a new harmonizing framework called HarmoFL. First, we propose to mitigate the local update drift by normalizing amplitudes of images transformed into the frequency domain to mimic a unified imaging setting, in order to generate a harmonized feature space across local clients. Second, based on harmonized features, we design a client weight perturbation guiding each local model to reach a flat optimum, where a neighborhood area of the local optimal solution has a uniformly low loss. Without any extra communication cost, the perturbation assists the global model to optimize towards a converged optimal solution by aggregating several local flat optima. We have theoretically analyzed the proposed method and empirically conducted extensive experiments on three medical image classification and segmentation tasks, showing that HarmoFL outperforms a set of recent state-of-the-art methods with promising convergence behavior. \mbox{Code is available at: \textcolor{blue}{\url{https://github.com/med-air/HarmoFL}}}
%Compared with classical FedAvg, as well as several state-of-the-art FL methods, HarmoFL demonstrates significant improvements in our extensive experiments. These empirical results are supported by a non-iid drift bounding analysis in a simplified setting.
\end{abstract}

\section{Introduction}

\begin{figure}[t]
\centering
\includegraphics[width=0.99\columnwidth]{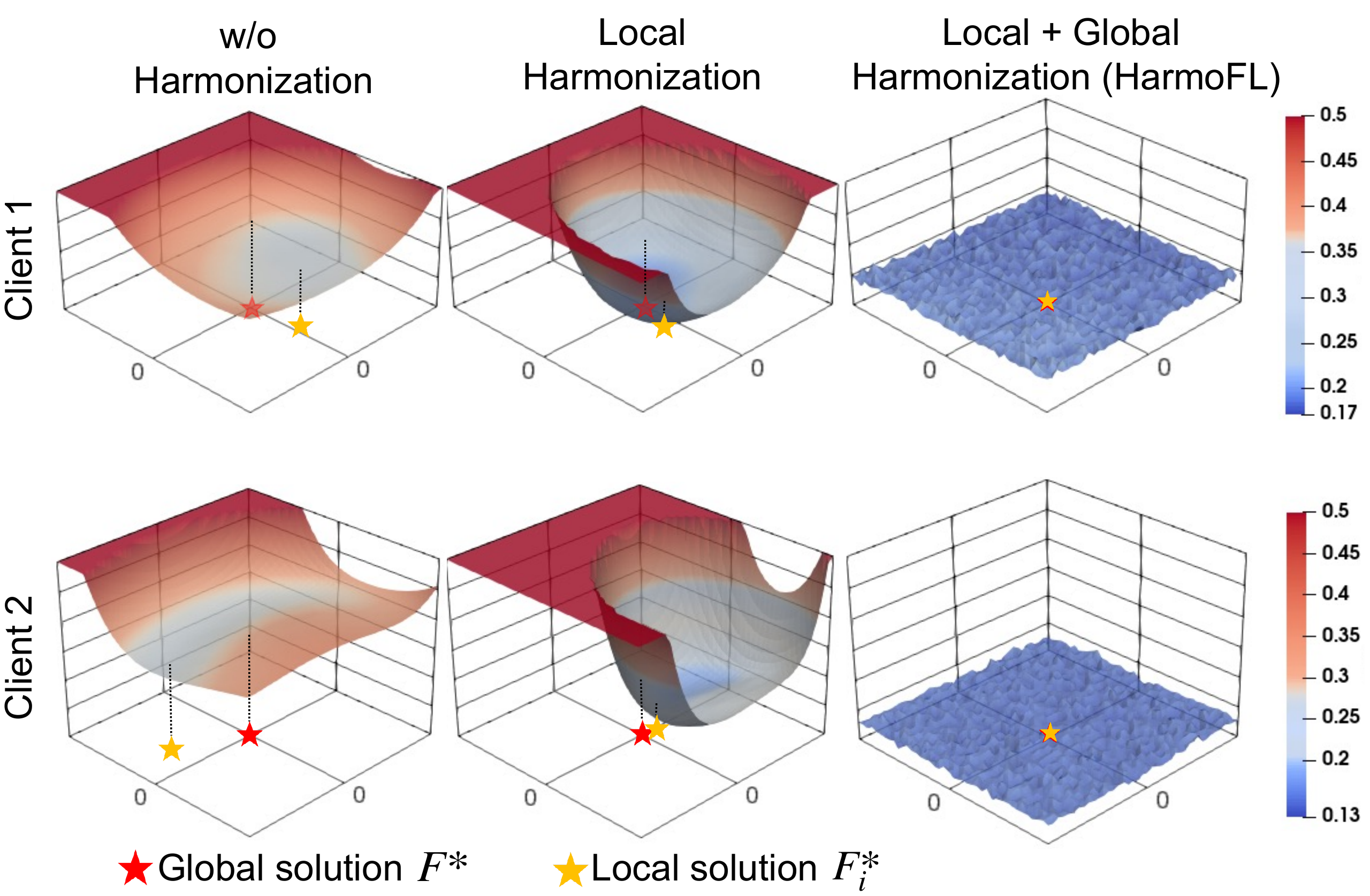} % Reduce the figure size so that it is slightly narrower than the column. Don't use precise values for figure width.This setup will avoid overfull boxes.
\vspace{-3mm}
\caption{Loss landscape visualization of two clients without harmonization (left), with only local harmonization (middle) and our complete method of HarmoFL (right). The vertical axis shows the loss (denoting the solution for global objective as $F^*$ and each local objective as $F_i^*$), 
and the horizontal plane represents a parameter space centered at the global model weight. (\emph{See Introduction for detailed explanation.})} 
\label{fig:cover_page}
\end{figure}

Multi-site collaborative training of deep networks is increasingly important for maximizing the potential of data-driven models in medical image analysis~\cite{shilo2020axes,peiffer2020machine,dhruva2020aggregating}, however, the data sharing is still restricted by some legal and ethical issues for protecting patient data. Federated learning recently allows a promising decentralized privacy-preserving solution, in which different institutions can jointly train a model without actual data sharing, i.e., training client models locally and aggregating them globally ~\cite{mcmahan2017communication,kaissis2020secure}.  

Despite the recent promising progress achieved by FL in medical image analysis~\cite{dou2021federated,rieke2020future,sheller2020federated,roth2020federated,ju2020federated}, the non-independent and identically distributed (non-iid) data is still an outstanding challenge in real-world practice~\cite{kairouz2019advances,hsieh2020non,xu2021federated}.
Non-iid issue typically happens, and the device vendors or data acquisition protocols are responsible for heterogeneity in the feature distributions~\cite{aubreville2020completely,liu2020ms}. For example, the appearance of histology images varies due to different staining situations, and MRI data of different hospitals suffer from feature distribution shifts associated with various scanners or imaging protocols.

Previous literature has demonstrated, both empirically and theoretically, that such data heterogeneity across clients introduces \emph{drift} in both local (client) and global (server) optimizations, making the convergence slow and unstable~\cite{zhao2018federated,li2019convergence,pmlr-v119-karimireddy20a}.
Specifically, in the local update, each client model will be optimized towards its own local optima (i.e., fitting its individual feature distribution) instead of solving the global objective, which raises a drift across client updates. 
Meanwhile, in the global update that aggregates these diverged local models, the server model is further distracted by the set of mismatching local optima, which subsequently leads to a global drift at the server model.
Fig.~\ref{fig:cover_page} intuitively illustrates such local and global drifts via loss landscape visualization~\cite{visualloss}, in an example of two non-iid clients.
The vertical axis shows the loss at each client (denoting the solution for global objective as $F^*$ and for each local objective as $F_i^*$), and the horizontal plane represents a parameter space centered at the specific parameters of global model . 
With the same objective function and parameter initialization for each client, we can see the local solution of two clients in the first column are significantly different, which indicates the drift across local client updates.
Globally, a current good solution $F^*$ may achieve a relatively low loss for both clients. However, since each client has its own shaped loss landscape, optimizing the current solution towards global optima is difficult and the aggregation of the two diverged local solutions further distracts the current good solution.
% since $F^*$ does not match any local optimal solution, 
% The  the local updates keeping towards $F_i^*$ will drive the global model away from the current solution and along a direction with high errors on the second client.

% From the first column, we can see that the local optima of two clients appear at different locations, indicating the client update drift. The optimal global model $F^*$ positioned at $(0,0)$ suffers from a high loss and shows a distance to reach any $F_i^*$ because of the server update drift. 

To address such non-iid problem in FL, existing works can be mainly divided into two groups, which correspond to tackling the drift locally or globally.
% \cite{sheller2018multi,fedsim,yeganeh2020inverse,fedbn}. 
The local side emphasizes how to better normalize those diverse data. For example, \cite{sheller2018multi} conducted a pioneer study and proposed using data pre-processing to reduce data heterogeneity. Recently, FedBN~\cite{fedbn} kept batch normalization layers locally to normalize local data distribution. 
As for the global server side, adjusting aggregation weight is a typical strategy, for instance, \cite{yeganeh2020inverse} used inverse distance to adaptively reweight aggregation. Recently, the framework FedAdam~\cite{fedadam} is proposed for introducing adaptive optimization to stabilize server update in heterogeneous data. However, these methods tackle heterogeneity partially via either client or server update. The local and global drifts are essentially coupled, yet how to jointly solve them as a whole still remains unclear. 

In this paper, we consider both client and server updates and propose a new harmonizing strategy to effectively solve the data heterogeneity problem for federated medical image analysis.
First, to mitigate local drift at the local update, we propose a more effective normalization strategy via amplitude normalization, which unifies amplitude components from images decomposed in the frequency domain. 
% With the amplitude normalization, non-iid features are harmonized into a similar space while structures are well preserved like images are generated and processed by a unified protocol. 
Then for the global drift, although weighted parameter aggregation is widely adopted, the coordination can fail over heterogeneous clients with large parameter differences. Thus, we aim to promote client models that are easy to aggregate rather than design a new re-weighting strategy. Based on the harmonized feature space, we design a weight-perturbation strategy to reduce the server update drift. The perturbation is generated locally from gradients and applied on the client model to constrain a neighborhood area of the local converged model to have a uniformly low loss. With the perturbation, each client finds a shared flat optimal solution that can be directly aggregated with others, assisting the global model to optimize towards a converged optimal solution. As can be observed from the last two columns of Fig.~\ref{fig:cover_page}, the proposed amplitude normalization for local harmonization well mitigates the distance between global and local solutions, and with the weight perturbation for global harmonization, our approach achieves flat client solutions that can be directly aggregated to obtain an optimal global model. 
% reaching almost the same position as local solutions, making all clients benefited from the federated learning. 

Our main contributions are highlighted as follows:
\begin{itemize}
\item We propose to effectively mitigate the local update drift by normalizing the frequency-space amplitude component of different images into a unified space, which harmonizes non-iid features across clients.

\item Based on the harmonized features, we further design a novel weight-perturbation based strategy to rectify global server update shift without extra communication cost.

\item To the best of our knowledge, we are the first to simultaneously address both local and global update drifts for federated learning on heterogeneous medical images. We have also theoretically analyzed the proposed HarmoFL framework from the aspect of gradients similarity, showing that drift caused by data heterogeneity is bounded.

\item We conduct extensive experiments on three medical image tasks, including breast cancer histology image classification, histology nuclei segmentation, and prostate MRI segmentation. Our HarmoFL significantly outperforms a set of latest state-of-the-art FL methods.
\end{itemize}

% In our extensive experiments, HarmoFL outperforms previous state-of-the-art on three different medical tasks and only requires minimal additional communication cost. Besides the superior performance, we also show a bounded drift between local and global optimal solutions by theoretically analyzing the optimization progress in a simplified setting.

% \begin{figure*}[t]
% \centering
% \includegraphics[width=0.99\textwidth]{pics/overview.pdf} % Reduce the figure size so that it is slightly narrower than the column.
% \caption{Not finished yet, many revisions need to be done.}
% \label{fig2}
% \end{figure*}

\section{Related Work}
There have been many methods proposed trying to improve federated learning on heterogeneous data and can be mainly divided into two groups, including improvements on local client training and on global server aggregation.
\paragraph{Local client training:} Literature towards improving the client training to tackle local drift includes, using data pre-processing to reduce data heterogeneity~\cite{sheller2018multi},
introducing a domain loss for heterogeneous electroencephalography classification~\cite{gao2019hhhfl}, using simulated CT volumes to mitigate scanner differences in cardiac segmentation~\cite{fedsim}, adding a proximal term to penalize client update towards smaller weight differences between client and server model~\cite{fedprox}, and learning affine transformation against shift under the assumption that data follows an affine distribution~\cite{fedrobust}. Recently, both SiloBN~\cite{silobn} and FedBN~\cite{fedbn} propose keeping batch normalization locally to help clients obtain similar feature distributions, and MOON~\cite{moon} uses contrastive learning on latent feature representations at the client-side to enhance the agreements between local and global models.
\paragraph{Global server aggregation:}
Besides designs on the client-side, many methods are proposed towards reducing global drift with a focus on the server-side. \cite{zhao2018federated} create a small subset of data that is globally shared across clients, and many other methods improve the aggregation strategy, e.g. \cite{yeganeh2020inverse} calculate the inverse distance to re-weight aggregation and FedNova~\cite{fednova} proposes to use normalized stochastic gradients to perform global model aggregation rather than the cumulative raw local gradient changes. Very recently, FedAdam~\cite{fedadam} introduces adaptive optimization into federated learning to stabilize convergence in heterogeneous data. 

However, the above-mentioned methods partially address the drift either from the local client or global server perspective. Instead, our approach aims jointly mitigating the two coupled drifts both locally and globally.

\section{Methodology}
To address the non-iid issues from both local and global aspects, we propose an effective new federated learning framework of HarmoFL. We start with the formulation of federated heterogeneous medical images analysis, then describe amplitude normalization and weight perturbation towards reducing local drift and global drift respectively. At last, we give a theoretical analysis of HarmoFL.
\subsection{Preliminaries}
Denote $(\X,\Y)$ as the joint image and label space over $N$ clients. A data sample is an image-label pair $(x,y)$ with $x \in \X, y \in \Y$, and the data sampled from a specific $i$-th client follows data distribution $\D_i$. In this work, we focus on the non-iid feature shift. Given the joint probability $P(\x,y)$ of feature $\x$ of image $x$ and label $y$, we have $P_i(\x)$ varies even if $P(y|\x)$ is the same or $P_i(\x|y)$ varies across clients while $P(y)$ is unchanged. 

Our proposed HarmoFL aims to improve federated learning for both local client training and global aggregation. The federated optimization objective is formulated as follows:
\begin{equation}
\label{eq:new_obj}
\min_{\theta} \left[F(\theta) := \sum_{i=1}^{N}p_i F_i(\theta + \delta, \overline{\D_i})\right],
\end{equation}
where $F_i = \sum_{(x,y) \sim \D_i}  \ell_i(\Psi(x),y;\theta + \delta)$ is the local objective function and $\ell_i$ is the loss function defined by the learned model $\theta$ and sampled pair $(x,y)$. For the $i$-th client, $p_i$ is the corresponding weight such that $p_i\ge0$ and $\sum_{i=1}^N p_i=1$. The $\delta$ is a weight perturbation term and $\overline{\D_i}$ is the harmonized feature distribution obtained by our amplitude normalization. Specifically, we propose a new amplitude normalization operator $\Psi(\cdot)$ harmonizing various client features distribution to mitigate client update drift. The normalizer manipulates amplitudes of data in the frequency space without violating the local original data preserving. Based on the harmonized features, we can generate a weight perturbation for each client without any extra communication cost. The perturbation forces each client reaching a uniformly low error in a neighborhood area of local optima, thus reducing the drift in the server. 

\begin{figure}[t]
\centering
\includegraphics[width=0.99\columnwidth]{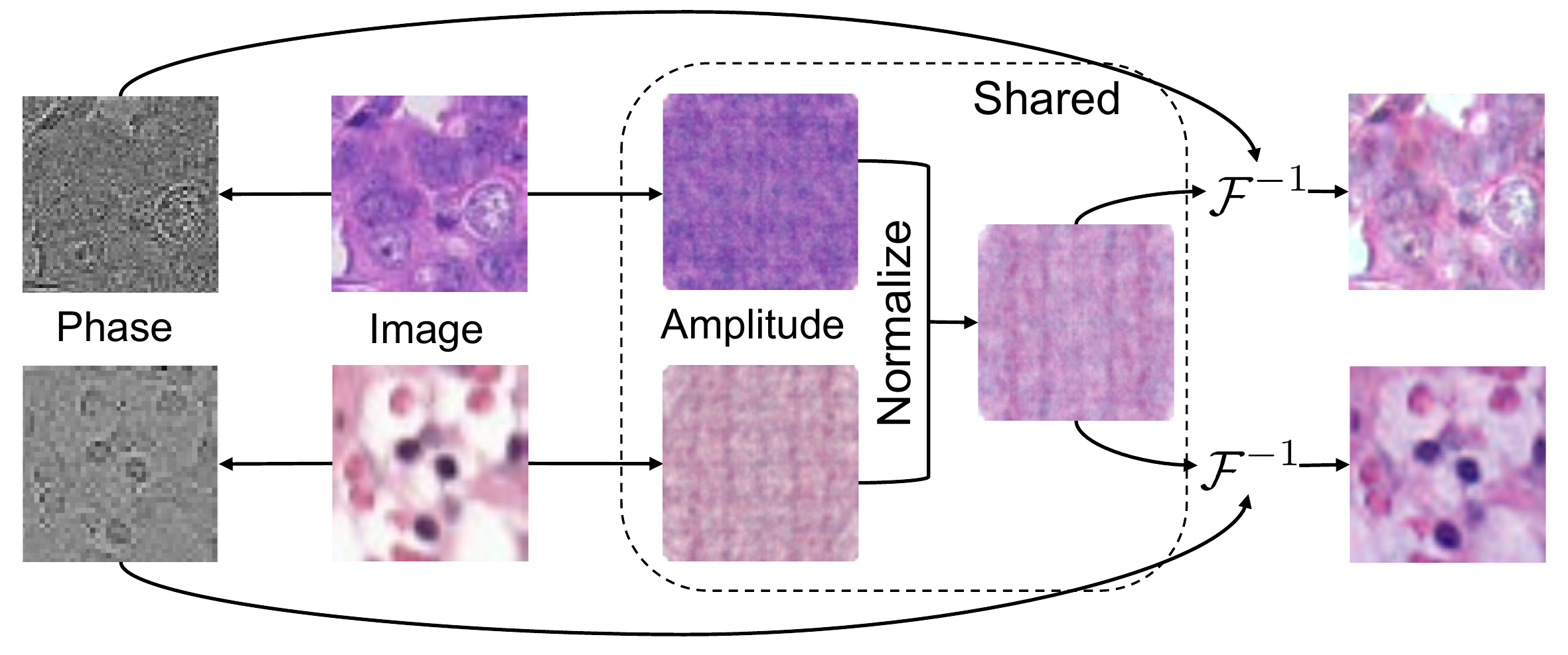} % Reduce the figure size so that it is slightly narrower than the column. Don't use precise values for figure width.This setup will avoid overfull boxes.
\vspace{-3mm}
\caption{Amplitude normalization that harmonizes local client features. Phase components are strictly kept locally and only the average amplitude from each client is shared.}
\label{fig:amp_norm}
\end{figure}

\subsection{Amplitude normalization for local training} 
Structure semantic is an important factor to support medical imaging analysis while low-level statistics (e.g. color, contrast) usually help to differentiate the structure. By decomposing the two parts, we are able to harmonize low-level features across hospitals while preserving critical structures. Using the fast Fourier transform~\cite{nussbaumer1981fast}, we can transform images into the frequency space signals and decompose out the amplitude spectrum which captures low-level features. Considering the patient level privacy sensitivity and forbiddance of sharing original images, we propose only using averaged amplitude term during communication to conduct the normalization. 

% So inspired from the fourier transformation, we can use fast fourier transform~\cite{nussbaumer1981fast} to get frequency space signals from original images and decompose out the amplitude spectrum and phase spectrum, which reflects low-level visual features and structure features. Then we normalize the amplitude in frequency space to harmonize non-iid features. 
More specifically, for input image $x$ from the \mbox{$i$-th} client, we transform each channel of image $x \in \R^{H\times W \times C}$ into the frequency space signals $\F_i(x)(u, v)=\sum_{h=0}^{H-1} \sum_{w=0}^{W-1} x(h, w) e^{-j 2 \pi\left(\frac{h}{H} u+\frac{w}{W} v\right)}$. Then we can split the real part $\mathcal{R}_i(x)$ and the imaginary part $\mathcal{I}_i(x)$ from the frequency signals $\F_i(x)$. The amplitude and phase component can be expressed as:
\begin{equation}
\label{eq:amp_phase}
\begin{gathered}
\mathcal{A}_i(x)=\left[\mathcal{R}_{i}^{2}(x)+\mathcal{I}_{i}^{2}(x)\right]^{{1}/{2}}\\ 
\mathcal{P}_i(x)=\arctan \left[{\mathcal{I}_i(x)}/{\mathcal{R}_i(x)}\right].
\end{gathered}
\end{equation}
Next, we normalize the amplitude component of each image batch-wise with a moving average term. For the $k$-th batch of $M$ sampled images, first we decompose the amplitude $\A_{i,x_m}$ and phase $\cP_{i,x_m}$ for each single image $x_m$ in the batch. 
% $$\begin{array}{l}\\ \mathcal{P}(x)(u, v)=\arctan \left[\frac{I(x)(u, v)}{R(x)(u, v)}\right]\end{array}$$
We calculate the in-batch average amplitude and update the average amplitude with a decay factor $v$:
\begin{equation}
\label{eq:amp_avg}
\overline{\A}_{i,k} = (1-v)\overline{\A}_{i,k-1} + v \frac{1}{M}\sum_{m=1}^M \A_{i,x_m},
\end{equation}
where $\overline{\A}_{i,k-1}$ is the average amplitude calculated from the previous batch and this term is set to zero for the first batch. The $\overline{\A}_{i,k}$ harmonizes low-level distributions inside a client and keeps tracking amplitudes of the whole client images. With the updated average term, each original image of the $k$-th batch is normalized using the average amplitude $\overline{\A}_{i,k}$ and its original phase component $\cP_{i,x_m}$ in below: 
\begin{equation} 
\label{eq:amp_norm}
\Psi(x_m) = \F^{-1}(\overline{\A}_{i,k}, \cP_{i,x_m}),
\end{equation}
where $\F^{-1}$ is the inverse Fourier transform. After client training, the amplitude normalization layer which only contains the average amplitude information will be sent to the central server and generate a global amplitude. This global amplitude compromises various low-level visual features across clients and can help reduce client update drift at the next federated round. In practice, we find that only communication at the first round and fix the global amplitude can well harmonize non-iid features as well as saving the communication cost. 
% As shown in the figure, the amplitude reconstruction only show color distributions, will not violate the data and patient privacy.

\subsection{Weight perturbation for global aggregation}
The above-proposed amplitude normalization allows each client to optimize within a harmonized feature space, mitigating the local drift. Based on the harmonized feature space, we further aim to promote client models that are easy to aggregate, thus rectifying the global drift. As the global aggregation is typically the weighted averaging, a client with flat optima can well coordinate with other clients than sharp local minima~\cite{keskar2016large}, which shows large error increases even with minor parameter change. 
Motivated from adversarial training, we propose a local optimization objective of each client as below:

\begin{equation}
\label{eq:adversarial_loss}
\min_{\theta} \sum_{(x,y)\sim\D_i} \max _{\left\|{x}^{\prime}-{x}\right\|_{p} \leq \delta} \ell({x}^{\prime},y;\theta),
\end{equation}
where $x^\prime$ is the adversarial image within a $L_p$-norm bounded $\delta$-ball centered at original image $x$, the adversarial image is generated with the same label of $x$ but different feature shift. However, the generating process has an extra communication burden since we have to transfer feature distribution information across clients. Based on the harmonized features, we carefully design a weight perturbation to effectively solve the Eq.~(\ref{eq:adversarial_loss}). 
With a little abuse of notation, instead of generating adversarial images bonded by the term $\delta$, we propose a new $\delta$ as a perturbation that is directly applied to model parameters. The perturbation is self-generated using gradients from the harmonized feature and has no extra communication cost. Formally, for the $k$-th batch, we first calculate the gradients of client model $\theta_{i,k-1}$ on the amplitude normalized feature $\Psi(x)$, which is calculated from Eq.~(\ref{eq:amp_norm}). Then we use the Euclidean norm $\| \cdot \|_2$ to normalize the gradients and obtain the perturbation term for the current iteration of gradient descent:
\begin{equation}
\label{eq:perturbation_term}
\delta_k =\alpha \frac{\nabla \ell_i(\Psi(x),y;\theta_{i,k-1})}{\left\|\ell_i(\Psi(x),y;\theta_{i,k-1})\right\|_{2}},
\end{equation}
where $\alpha$ is a hyper-parameter to control the degree of perturbation. The flat area around the local client optimum can be expanded with the larger $\alpha$, 
and the choice of $\alpha$ is studied in the experiment section.
After obtaining the perturbation term $\delta_k$, we minimize the loss on the parameter-perturbated model as below:
\begin{equation}
\label{eq:perturb_update}
\theta_{i,k} \leftarrow \theta_{i,k-1} - \eta_l \nabla \ell_i(\Psi(x),y;\theta_{i,k-1} + \delta_k),
\end{equation}
where $\theta_{i,k-1}$ is the model from the previous batch and $\eta_l$ is the local client learning rate. After iteratively update, each local client model is gradually driven towards an optimal solution that holds a uniformly low loss around its neighborhood area (i.e. flat optimum), thus promoting the aggregation and getting rid of trapped into the local sharp minima. 
% flat optima. The perturbation based optimization constrains each local client to findThe drift in global server aggregation is well reduced since local clients share more weight space overlaps with low errors.

\begin{algorithm}[tb]
\caption{Harmonizing Local and Global Drifts in FL}
\label{alg:algorithm}
\textbf{Input}: communication rounds $T$, number of clients $N$, mini-batch steps $K$, client learning rate $\eta_l$, global learning rate $\eta_g$, hyper-parameter $\alpha$ \\
% \textbf{Parameter}: Optional list of parameters\\
\textbf{Output}: The final global model $\theta^{(T)}$
\begin{algorithmic}[1] %[1] enables line numbers
\State Initialize server model $\theta^{(1)}$ 
% and amplitude normalizer $\overline{\A}^{(0)}$
\For{$t=1,2,\cdots, T$}
\For{$i=1,2,\cdots, N$ \textbf{in parallel}} 
\State  $\theta_{i,1}^t \leftarrow \theta^t$ \Comment{send the global model $\theta^t$ to client $i$}
% \State $\overline{\A}_{i,0}^t \leftarrow 0$
\For{$k=1,2,\cdots,K$} \Comment{client training}
\State sample a batch of data pairs $(x,y)$ of $\D_i$ 
% \If{t $\le$ 2}:
\State $\overline{\A}_{i,k}^t = (1- v)\overline{\A}_{i,k-1}^t$
\State $\quad\quad\quad + v \frac{1}{M}\sum_{m=1}^M\A_{x_m}$ \Comment{Eq.~(\ref{eq:amp_avg})}
% \Else
% \State $\overline{\A}_{i,k}^t = \overline{\A}^{t}$
% \EndIf
\State $\Psi(x) = \F^{-1}(\overline{\A}_{i,k}^t,\cP_{x_m})$ \Comment{Eq.~(\ref{eq:amp_norm})}
\State $\delta_k =\alpha \frac{\nabla \ell_i(\Psi(x),y;\theta)}{\left\|\ell_i(\Psi(x),y;\theta)\right\|_{2}}$ \Comment{Eq.~(\ref{eq:perturbation_term})}
\State $g_\delta = \nabla \ell_i(\Psi(x),y;\theta_i^t + \delta_k)$
\State $\theta_{i,k+1}^t \leftarrow \theta_{i,k}^t - \eta_l g_\delta$  \Comment{Eq.~(\ref{eq:perturb_update})}
\EndFor
\State \textbf{return} $\theta_{i,K}^t$ \Comment{send client model to server}
\EndFor
\State $\theta^{t+1} \leftarrow \theta^t - \eta_g \sum_{i=1}^N p_i (\theta_{i,K}^t - \theta_{i,1}^t)$ 
\State $\overline{\A}^{t+1} = \frac{1}{N}\sum_{i=1}^N \overline{\A}_{i,K}^t$ 
\EndFor 
\State \textbf{return} $\theta^{(T)}$
\end{algorithmic}
\end{algorithm}

\subsection{Theoretical analysis for HarmoFL}
With the help of amplitude normalization and weight perturbation, we constrain the client update with less dissimilarity and achieve a model having a relatively low error with a range of parameter changes.
To theoretically analyze HarmoFL, we transform our approach into a simplified setting where we interpret both the amplitude normalization and weight perturbation from the gradients perspective.
For amplitude normalization, it reduces gradients dissimilarity between clients by harmonizing the non-iid features.
In the meanwhile, the weight perturbation forces clients to achieve flat optima, in which the loss variation is constrained, making the gradients change mildly. So both amplitude normalization and weight perturbation bound gradient differences across clients, and this assumption has also been widely explored in different forms~\cite{yin2018gradient,fedprox,vaswani2019fast,karimireddy2019scaffold}. Based on the standard federated optimization objective, we formulate a new form in below:
% to guide a consist optimization direction of each client. 
\begin{equation}
\label{eq:grad_constrain}
\begin{gathered}
% \begin{aligned}
\min_{\theta} \left[F(\theta) := \sum_{i=1}^{N}p_i F_i(\theta, \D_i) \right] \\ 
s.t.\sum_{\substack{(x_i,y_i)\sim \D_i,\\ (x_j,y_j)\sim \D_j}} |\nabla \ell(x_i,y_i;\theta) - \nabla \ell(x_j,y_j;\theta)| \le \epsilon,
\end{gathered}
\end{equation}
% \end{aligned}
% \in \{1,\cdots,m\} \backslash \{i\}
where $j \in \{1,\cdots,N\} \backslash \{i\} $ and $\epsilon$ is a non-negative constant and $(x_i,y_i)$ is the image-label pair from the $i$-th client's distribution. To quantify the overall drift between client and server models at the $t$-th communication round, we define the overall drift term as below:
\begin{equation}
\label{eq:overall_shift}
\Gamma=\frac{1}{KN}\sum_{k=1}^K \sum_{i=1}^N \E [ \|\theta_{i,k}^t - \theta^t \|^2 ],
\end{equation}
where $K$ is the mini-batch steps. We theoretically show that with the bounded gradient difference, our HarmoFL strategy is guaranteed to have an upper bound on the overall drift caused by non-iid data.
\begin{theorem}
\label{thm:bounded_drift}
With the shift term $\Gamma$ defined in Eq.~(\ref{eq:overall_shift}), assume the gradient dissimilarity and variance are bounded and the functions $F_i$ are $\beta$-smooth, denote the effective step-size $\tilde{\eta}=K\eta_g\eta_l$, we have the upper bound for the overall drift of our HarmoFL below:
\begin{align*}
\Gamma &\leq \frac{1}{N}\sum_{i=1}^N \frac{4\tilde{\eta}^2}{\eta_g^2}\|\nabla F_i(\theta)\|^2 + \frac{4\tilde{\eta}^2\epsilon^2(N-1)^2}{\eta_g^2 N^2} + \frac{2\tilde{\eta}^2\sigma^2}{K\eta_g^2}\\
&\leq \frac{4\tilde{\eta}^2(G^2+B^2C^2)}{\eta_g^2} + \frac{4\tilde{\eta}^2\epsilon^2(N-1)^2}{\eta_g^2 N^2} + \frac{2\tilde{\eta}^2\sigma^2}{K\eta_g^2},
% \frac{4\tilde{\eta}^2B^2}{\eta_g^2}\|\nabla F(\theta)\|^2
\end{align*}
\end{theorem}
where $C:=F(\theta^0) - F^*$ when functions $F_i$ are convex, and for non-convex situation, we have $C:=\|\nabla F(\theta)\|$. This theorem gives the upper bound with both the convex and non-convex assumptions for $F_i$.

Please find the notation table in Appendix~\ref{app:notation_table}. All assumptions and proofs are formally given in Appendix~\ref{app:proof}. The proof sketch is applying our extra gradient differences constraint in the subproblem of one round optimization and using the bounded dissimilarity. The shift bound is finally unrolled from a recursion format of one round progress.

% \subsection{Implementation details}0

\section{Experiments}
% Please add the following required packages to your document preamble:
In this section, we extensively evaluate our method to demonstrate that harmonizing local and global drifts are beneficial for clients with heterogeneous features. Our harmonizing strategy, HarmoFL, achieves higher performance as well as more stable convergence compared with other methods on feature non-iid datasets. This is shown on the breast cancer histology image classification, histology nuclei segmentation, and prostate MRI segmentation. All results reported are the average of three repeating runs with a standard deviation of different random seeds. More results please refer to Appendix~\ref{app:complete_exp}.

\subsection{Dataset and experimental settings}
\begin{figure}[h]
\centering
\includegraphics[width=0.99\columnwidth]{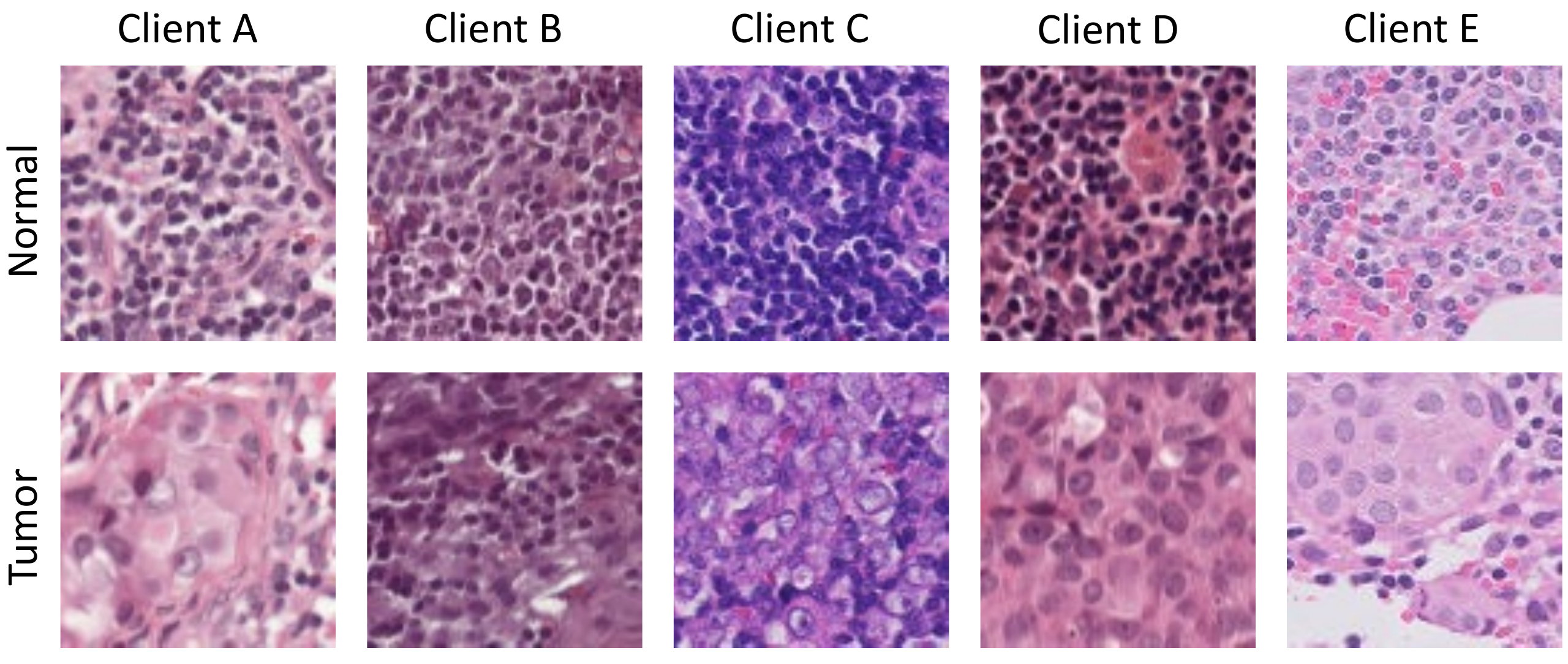} 
\vspace{-3mm}
\caption{Examples of breast histology images of normal and tumor tissues from five clients, showing large heterogeneity.}
\label{fig:histo_samples}
\end{figure}

\paragraph{Breast cancer histology image classification.} We use the public tumor dataset Camelyon17, which contains 450,000 histology images with different stains from 5 different hospitals~\cite{bandi2018detection}. As shown in Fig.~\ref{fig:histo_samples}, we take each hospital as a single client, and images from different clients have heterogeneous appearances but share the same label distribution (i.e. normal and tumor tissues). We use a deep network of DenseNet121~\cite{huang2017densely} and train the model for 100 epochs at the client-side with different communication frequencies. We use cross-entropy loss and SGD optimizer with a learning rate of $1e^{-3}$.

\paragraph{Histology nuclei segmentation.} For the cell nuclei segmentation, we gather three public datasets, including MoNuSAC2020~\cite{verma2021monusac2020}, MoNuSAC2018~\cite{kumar2019multi} and TNBC~\cite{naylor2018segmentation}. For data from MoNuSAC2020, we divide them into 4 clients according to different hospitals they come from and form 6 clients in total.
We use U-Net~\cite{ronneberger2015u} and train the model for 500 communication rounds with 1 local update epoch for each communication round.
We use segmentation Dice loss and Adam optimizer with learning rate of $1e^{-4}$, momentum of 0.9 and 0.99.

\paragraph{Prostate MRI segmentation.}
For the prostate segmentation, we use a multi-site prostate segmentation dataset~\cite{liu2020ms} which contains 6 different data sources from 3 public datasets \cite{nciisbi2013,lemaitre2015computer,litjens2014evaluation}. We regard each data source as a client and train the U-Net using Adam optimizer with a learning rate of $1e^{-4}$, momentum of 0.9 and 0.99.

% For all clients in three datasets, we use 20\% of client data for test, for the reaming data, we use 80\% for train and 20\% for validation.
We report the performance of global models, i.e., the final results of our overall framework. The model was selected using the separate validation set and evaluated on the testing set. To avoid distracting focus on the feature non-iid problem due to data imbalance, we truncate the sample size of each client to their respective smaller number in histology image classification and prostate MRI segmentation task, but we keep the client data imbalance in the nuclei segmentation task to demonstrate the performance with data quantity difference. If not specified, our default setting for the local update epoch is $1$. We use the momentum of $0.9$ and weight decay of $1e^{-4}$ for all optimizers in three tasks. We empirically set the decay factor $v$ to $0.1$ and set the degree of perturbation term to $5e^{-2}$ by grid search. For more dataset and implementation details, please refer to Appendix~\ref{app:complete_exp}.

\begin{table*}[t]
\centering
% \small
\setlength\tabcolsep{3.5pt}
\caption{Results for histology nuclei segmentation and prostate MRI segmentation. The results of the Dice coefficient are reported. Each column represents one client and the Avg. is abbreviated for the average Dice.}
\vspace{-3mm}
\begin{tabular}{cccccccc||ccccccc}
\toprule
\multirow{2}{*}{Method} & \multicolumn{7}{c}{Histology Nuclei Segmentation (Dice \%)}
& \multicolumn{7}{c}{Prostate MRI Segmentation (Dice \%)}\\ \cline{2-15} 
& A & B & C & D & E & F & Avg. & A & B & C & D & E & F & Avg. \\ \hline \\[-1.8ex]

FedAvg & 73.44 & 73.06 & 72.52 & 68.91 & 67.33 & 49.69 & 67.49 & 90.04 & 94.31 & 92.60 & 92.21 & 90.14 & 89.36 & 91.44 \\ \\[-2.9ex]
(PMLR2017) & \small{(0.02)} & \small{(0.24)} & \small{(0.85)} & \small{(0.34)} & \small{(0.86)} & \small{(0.34)} & \small{(9.06)} & \small{(1.27)} & \small{(0.28)} & \small{(0.66)} & \small{(0.71)} & \small{(0.27)} & \small{(1.76)} & \small{(1.91)} \\ [0.4ex]

FedProx & 73.49 & 73.11 & 72.45 & 69.01 & 67.33 & 49.56 & 67.49 & 90.65 & 94.60 & 92.64 & 92.19 & 89.36 & 87.07 & 91.08 \\ \\[-2.9ex]
(MLSys2020) & \small{(0.07)} & \small{(0.19)} & \small{(0.94)} & \small{(0.34)} & \small{(0.86)} & \small{(0.34)} & \small{(9.12)} & \small{(1.95)} & \small{(0.30)} & \small{(1.03)} & \small{(0.15)} & \small{(0.97)} & \small{(1.53)} & \small{(2.66)} \\[0.4ex]

FedNova & 73.40 & 73.01 & 71.50 & 69.23 & 67.46 & 50.68 & 67.55 & 90.73 & 94.26 & 92.73 & 91.91 & 90.01 & 89.94 & 91.60 \\ \\[-2.9ex]
(NeurIPS2020) & \small{(0.05)} & \small{(0.38)} & \small{(1.35)} & \small{(0.34)} & \small{(1.07)} & \small{(0.34)} & \small{(8.57)} & \small{(0.41)} & \small{(0.08)} & \small{(1.29)} & \small{(0.61)} & \small{(0.87)} & \small{(1.54)} & \small{(1.70)} \\ [0.4ex]
FedAdam & 73.53 & 72.91 & 71.74 & 69.26 & 66.69 & 49.72 & 67.31 & 90.02 & 94.84 & 93.30 & 91.70 & 90.17 & 87.77 & 91.30 \\ \\[-2.9ex]
(ICLR2021) & \small{(0.08)}& \small{(0.24)} & \small{(1.33)} & \small{(0.50)} & \small{(1.18)} & \small{(0.11)} & \small{(8.98)} & \small{(0.29)} & \small{(0.11)} & \small{(0.79)} & \small{(0.16)} & \small{(1.46)} & \small{(1.35)} & \small{(2.53)} \\ [0.4ex]

FedBN & 72.50 & 72.51 & 74.25 & 64.84 & 68.39 & 69.11 & 70.27 & 92.68 & 94.83 & 93.77 & 92.32 & 93.20 & 89.68 & 92.75 \\ \\[-2.9ex]
(ICLR2021) & \small{(0.81)} & \small{(0.13)} & \small{(0.28)} & \small{(0.93)} & \small{(1.13)} & \small{(0.94)} & \small{(3.47)} & \small{(0.52)} & \small{(0.47)} & \small{(0.41)} & \small{(0.19)} & \small{(0.45)} & \small{(0.60)} & \small{(1.74)} \\ [0.4ex]

MOON & 72.85 & 71.92 & 69.23 & 69.00 & 65.08 & 48.26 & 66.06 & 91.79 & 93.63 & 93.01 & 92.61 & 91.22 & 91.14 & 92.23 \\ \\[-2.9ex]
(CVPR2021) & \small{(0.46)} & \small{(0.37)} & \small{(2.29)} & \small{(0.71)} & \small{(0.73)} & \small{(0.66)} & \small{(9.13)} & \small{(1.64)} & \small{(0.21)} & \small{(0.75)} & \small{(0.53)} & \small{(0.61)} & \small{(0.88)} & \small{(1.01)} \\ [0.4ex]

HarmoFL & \textbf{74.98} & \textbf{75.21} & \textbf{76.63} & \textbf{76.59} & \textbf{73.94} & \textbf{69.20} & \textbf{74.42} & \textbf{94.06} & \textbf{95.26} & \textbf{95.28} & \textbf{93.51} & \textbf{94.05} & \textbf{93.53} & \textbf{94.28} \\ \\[-2.9ex]
(\textbf{Ours}) & \small{\textbf{(0.36)}}  & \small{\textbf{(0.57)}}  & \small{\textbf{(0.20)}}  & \small{\textbf{(0.77)}}  & \small{\textbf{(0.13)}}  & \small{\textbf{(1.23)}}  & \small{\textbf{(2.76)}}  & \small{\textbf{(0.47)}}  & \small{\textbf{(0.38)}}  & \small{\textbf{(0.33)}}  & \small{\textbf{(0.79)}}  & \small{\textbf{(0.50)}}  & \small{\textbf{(1.02)}}  & \small{\textbf{(0.80)}} \\
\bottomrule
\end{tabular}
\label{table:seg_res}
\end{table*}

\begin{table}[ht]
\small
\centering
\setlength\tabcolsep{2.7pt}
\caption{Results for breast cancer histology images classification  of different methods. Each column represents one client and the Avg. is abbreviated for the average accuracy.}
\vspace{-3mm}
\begin{tabular}{ccccccc}
\toprule
\multirow{3}{*}{Method} & \multicolumn{6}{c}{Breast Cancer Histology Image Classification } \\
& \multicolumn{6}{c}{(Accuracy \%)}
\\ \cline{2-7} 
                        & A    & B    & C    & D    & E & Avg.\\ \hline \\[-1.8ex]
FedAvg & 91.10  & 83.12 & 82.06 & 87.49 & 74.78 & 83.71\\ \\[-2.9ex]
(PMLR2017) & (0.46)  & (1.58) & (8.52) & (2.49) & (3.19) & (6.16)\\
[0.4ex]
FedProx & 91.03 & 82.88 & 82.78 & 87.07 & 74.93 & 83.74 \\ \\[-2.9ex]
(MLSys2020) & (0.50) & (1.63) & (8.56) & (1.76) & (3.05) & (5.99) \\ 
[0.4ex]
FedNova & 90.99  & 82.97 & 82.40 & 86.93 & 74.86 & 83.61\\ \\[-2.9ex]
(NeurIPS2020) & (0.54)  & (1.76) & (9.21) & (1.58) & (3.12) & (6.00)\\
[0.4ex]
FedAdam & 87.45  & 80.38 & 76.89 & 89.27 & 77.86 & 82.37\\ \\[-2.9ex]
(ICLR2021) & (0.77)  & (2.03) & (14.03) & (1.28) & (2.68) & (5.65)\\
[0.4ex]
FedBN  & 89.35  & 90.25 & 94.16 & 94.04 & 68.87 & 87.33 \\ \\[-2.9ex]
(ICLR2021)  & (8.50) & (1.66) & (1.00) & (2.32) & (22.14) & (10.55) \\
[0.4ex]
MOON & 88.92 & 83.52 & 84.71 & 90.02 & 67.79 & 82.99 \\ \\[-2.9ex]
(CVPR2021) & (1.54) & (0.31) & (5.14) & (1.56) & (2.06) & (8.93) \\
[0.4ex]
HarmoFL &\textbf{ 96.17}  & \textbf{93.60} & \textbf{95.54} &\textbf{ 95.58} & \textbf{96.50} & \textbf{95.48} \\ \\[-2.9ex]
(\textbf{Ours}) &\textbf{(0.56)}  & \textbf{(0.67)} & \textbf{(0.32)} &\textbf{(0.27)} & \textbf{(0.46)} & \textbf{(1.13)} \\
\bottomrule
\end{tabular}
\label{table:breast_diagnosis}
\end{table}

\subsection{Comparison with the state-of-the-arts}
We compare our approach with recent state-of-the-art (SOTA) FL  methods towards solving the non-iid problem. For local drifts, \textbf{FedBN}~\cite{fedbn} focuses on the non-iid feature shift with medical image applications, and both \textbf{FedProx}~\cite{fedprox} and a recent method \textbf{MOON}~\cite{moon} tackle the non-iid problem by constraining the dissimilarity between local and global models to reduce global aggregation shifts. \textbf{FedAdam}~\cite{fedadam} and \textbf{FedNova}~\cite{fednova} are proposed as general methods to tackle global drifts. For the breast cancer histology image classification shown in Table~\ref{table:breast_diagnosis}, we report the testing accuracy on five different clients and the average results. FedProx only achieves minor improvements than FedAvg, showing that only reducing global aggregation drift may not achieve promising results when local clients shifted severely. Another recent representation dissimilarity constraining method, MOON, boosts the accuracy on clients B, C, D but suffers from a large drop on client E. The reason may come from that images in client E appear differently as shown in Fig.~\ref{fig:histo_samples}, making the representations of client E failed to be constrained towards other clients. With the harmonized strategy reducing both local and global drifts, our method consistently outperforms others, reaching an accuracy of 95.48\% on average, which is 8\% higher than the previous SOTA (FedBN) for the feature non-iid problem. Besides, our method can help all heterogeneous clients benefit from the federated learning, where clients show a testing accuracy with a small standard deviation on average, while other methods show larger variance across clients.

\begin{figure*}[t]
\centering
\includegraphics[width=1.9\columnwidth]{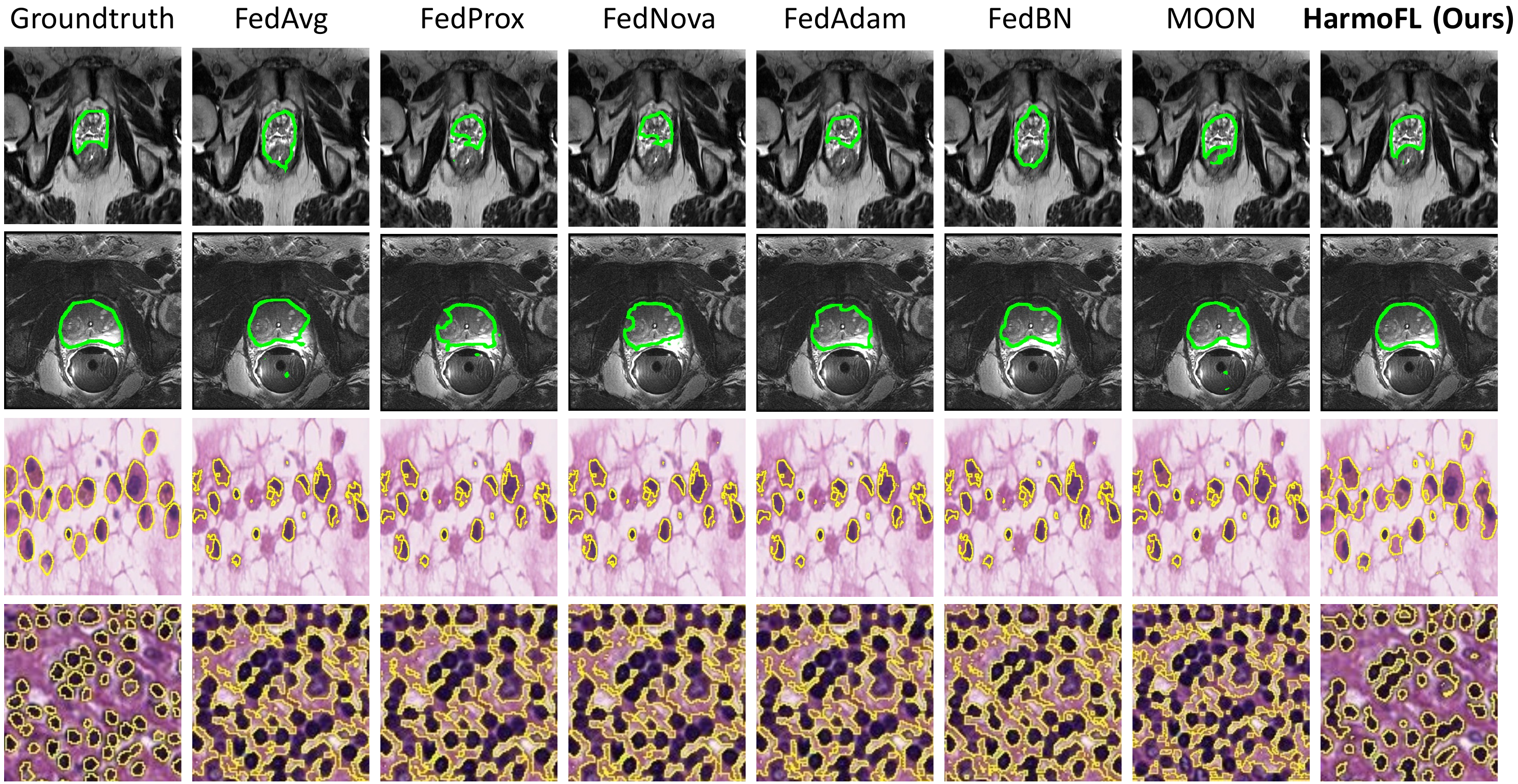} % Reduce the figure size so that it is slightly narrower than the column. Don't use precise values for figure width.This setup will avoid overfull boxes.
\vspace{-1mm}
\caption{Qualitative comparison on segmentation results with our method and other state-of-the-art methods. Top two rows for the task of prostate MRI segmentation and the bottom two rows for the task of histology nuclei segmentation.}
\label{fig:res_vis}
\vspace{-1mm}
\end{figure*}

For segmentation tasks, the experimental results of Dice are shown in Table~\ref{table:seg_res} in the form of single client and average performance. On histology nuclei segmentation, HarmoFL significantly outperforms the SOTA method of FedBN and improves at least 4\% on mean accuracy compared with all other methods. On the prostate segmentation task, as MRI images show fewer non-iid feature shifts than histology images, the performance gap of each client is not as large as nuclei segmentation. However, our method still consistently achieves the highest Dice of 94.28\% and has a smaller standard deviation over different clients. Besides, we visualize the segmentation results to demonstrate a qualitative comparison, as shown in Fig.~\ref{fig:res_vis}. Comparing with the first ground-truth column, due to the heterogeneous features, other federated learning methods either cover more or fewer areas in both prostate MRI and histology images. As can be observed from the second and fourth row, the heterogeneity in features also makes other methods fail to obtain an accurate boundary. But with the proposed harmonizing strategy, our approach shows more accurate and smooth boundaries.

% \begin{figure}[t]
% \centering
% \includegraphics[width=1\columnwidth]{pics/communication_rounds_smooth_wider.pdf}
% % \includegraphics[width=1\columnwidth]{pics/communication_plots_vertical.pdf} 
% \caption{The convergence of different methods in terms of testing accuracy with communicatino rounds.}
% % Our approach (HarmoFL) shows more higher and stable improvements on accuracy compared with other methods and all clients benefited from the federated learning.}
% \label{fig:communication_rounds}
% \end{figure}

\begin{figure}[t]
\centering
\includegraphics[width=0.9\columnwidth]{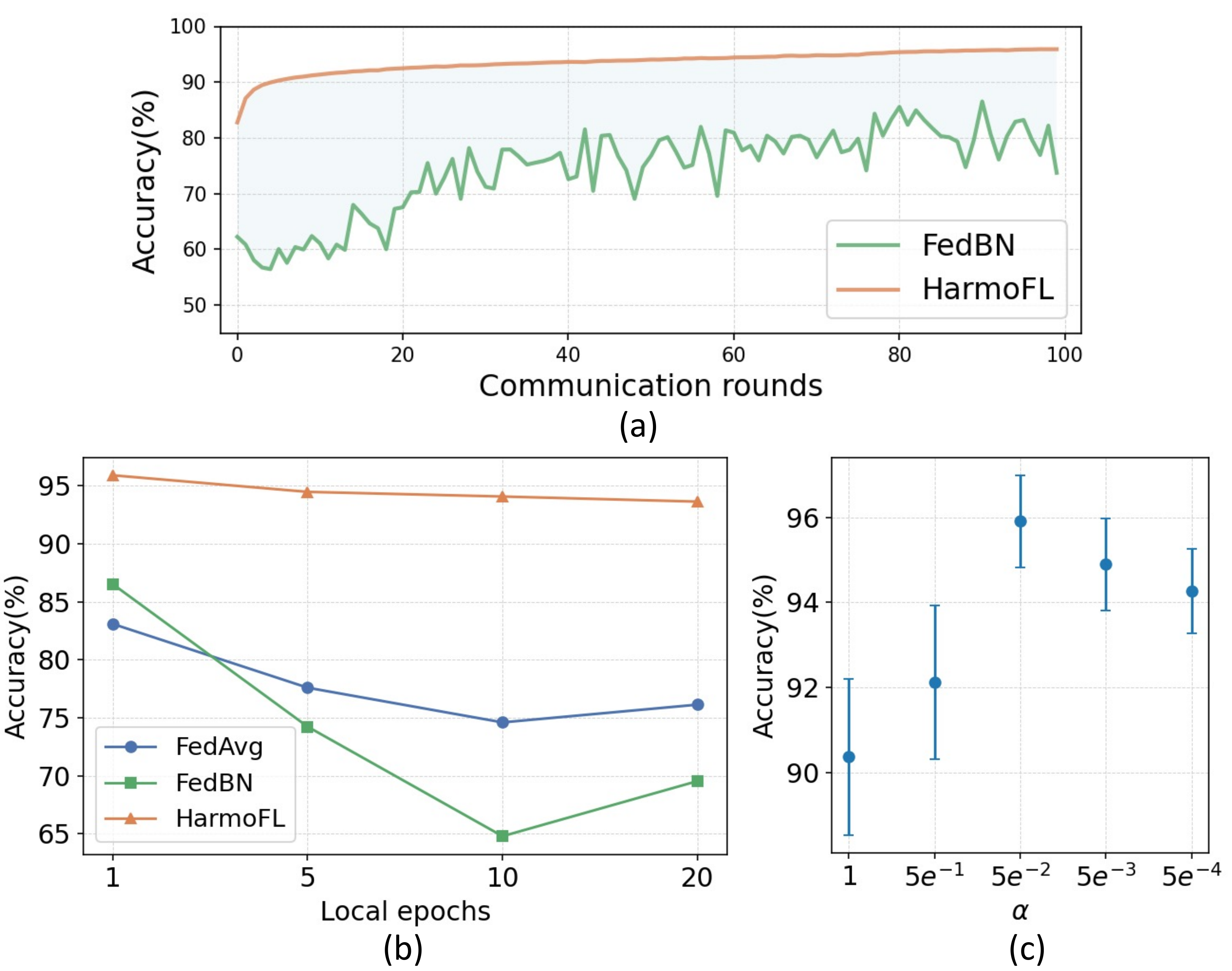} 
\vspace{-3mm}
\caption{(a) Convergence in terms of testing accuracy with communication rounds. (b) Comparison of FedAvg, FedBN, and our HarmoFL with different local training epochs. (c) Performance of HarmoFL with different perturbation radius.}
% The global model tends to degenerate with larger drifts caused by more local epochs, while HarmoFL behaves more robust to the local epoch change. 
\label{fig:ablation_study}
\vspace{-3mm}
\end{figure}

\subsection{Ablation study}
We further conduct ablation study based on the breast cancer histology image classification to investigate key properties of our HarmoFL, including the convergence analysis, influence of different local update epochs, and the effects of weight perturbation degree.
\\
\\\textbf{Convergence analysis.} To demonstrate the effectiveness of our method on reducing local update and server update drifts, we plot the testing accuracy curve on the average of five clients for $100$ communication rounds with $1$ local update epoch. As shown in Fig.~\ref{fig:ablation_study}(a), the curve of HarmoFL increases smoothly with communication rounds increasing, while the state-of-the-art method of FedBN~\cite{fedbn}, shows unstable convergence as well as lower accuracy. From the curve of FedBN~\cite{fedbn}, we can see the there are almost no improvements at the first 10 communication rounds, the potential reason is that the batch normalization layers in each local client model are not fully trained to normalize feature distributions. 
% and even inconsistent optimization directions. Specifically, the testing accuracy of Client A-D with FedBN increases more or less, but for Client E, the accuracy remains almost unchanged. This may due to the significant differences in features distracts the joint optimization direction of each client, making client competes with each other instead of jointly optimization towards a better global model. As for other federated learning methods, they even fail to converge at the very beginning of the training due to the large non-iid feature shifts.
\\\\\textbf{Influence of local update epochs.}
Aggregating with different frequencies may affect the learning behavior, since less frequent communication will further enhance the drift due to the non-iid feature, and finally obtaining a worse global model. We study the effectiveness of HarmoFL with different local update epochs and the results are shown in Fig.~\ref{fig:ablation_study}(b). When the local epoch is 1, each client communicates with others frequently, all methods have a relatively high testing accuracy. With more local epochs added, the local client update drift is increased and the differences between local and global models become larger. Both FedAvg and FedBN suffer from large drifts and show severe performance drops. However, because our method enables each client to train the model using weight perturbation with harmonized features, HarmoFL significantly outperforms other approaches and shows robustness to larger drift brought by more local update epochs.
\\\\
\textbf{Effects of weight perturbation degree.}
We further analyze how the weight perturbation degree control hyper-parameter $\alpha$ affects the performance of our method. Intuitively, the value of $\alpha$ indicates the radius of the flat optima area. 
A small radius may hinder clients to find a shared flat area during aggregation, while a large radius creates difficulties in the optimization of reaching such a flat optimum. As shown in Fig~\ref{fig:ablation_study}(c), we plot the average testing accuracy with standard error across clients by searching $\alpha \in \{1, 5e^{-1},5e^{-2},5e^{-3},5e^{-4}\}$. Our method reaches the highest accuracy when $\alpha=5e^{-2}$ and the performance decreases with $\alpha$ reducing. Besides, we can see our method can achieve more than 90\% accuracy even without degree control (i.e. $\alpha=1$).

\section{Conclusion}
This work proposes a novel harmonizing strategy, HarmoFL, which uses amplitude normalization and weight perturbation to tackle the drifts that exist in both local client and global server. Our solution gives inspiration on simultaneously solving the essentially coupled local and global drifts in FL, instead of regarding each drift as a separate issue. We conduct extensive experiments on heterogeneous medical images, including one classification task and two segmentation tasks, and demonstrates the effectiveness of our approach consistently. We further provide theoretical analysis to support the empirical results by showing the overall non-iid drift caused by data heterogeneity is bounded in our proposed HarmoFL. Overall, our work is beneficial to promote wider impact of FL in real world medical applications.

\bibliography{aaai22}
\newpage
\appendix
\paragraph{Roadmap of Appendix:} The Appendix is organized as follows. We list the notations table in Section~\ref{app:notation_table}. We provide formal assumptions, theoretical analysis, and proof of bounded non-iid drift in Section~\ref{app:proof}. The details of the experimental setting and additional results are in Section~\ref{app:complete_exp}.
\section{Notation Table}
\label{app:notation_table}
\begin{table}[ht]
\centering
% \resizebox{\textwidth}{!}{%
\begin{tabular}{cl}
\hline
Notations & Description \\ \hline \\[-1.8ex]
$F$ & global objective function. \\[0.3ex] 
$F_i$ & local objective function. \\[0.3ex] 
$(\X, \Y)$ & joint image and label space. \\[0.3ex]
$(x, y)$ &  a pair of data sample. \\[0.3ex]
$T$ & communication rounds.\\ [0.3ex]
$N$ & number of clients.\\ [0.3ex]
$K$ & mini-batch steps. \\[0.3ex]
$M$ & size of sampled images in one batch. \\[0.3ex]
\multirow{2}{*}{$\ell_i$} & loss function defined by the learned model \\ & and sampled pair. \\[0.3ex]
\multirow{2}{*}{$p_i$} & weight in the federated optimization \\ & objective of the $i$-th client. \\[0.3ex] 
$\D_i$ & data distribution for the $i$-th client.  \\ [0.3ex]
$\Psi(\cdot)$ & the amplitude normalization operation. \\[0.3ex]
\multirow{2}{*}{$\overline{\D_i}$} & distribution harmonized by \\ & amplitude normalization. \\ [0.3ex]
$\A_{i,x_m}$ & amplitude for each single image in a batch.\\[0.3ex]
$\cP_{i,x_m}$ & phase for each single image in a batch.\\[0.3ex]
\multirow{2}{*}{$\overline{\A}_{i,k}$} & average amplitude calculated from \\ & the current batch.\\ [0.3ex]
$\delta$ & perturbation applied to model parameters.\\ [0.3ex]
\multirow{2}{*}{$\alpha$} & hyper-parameter to control the degree \\ & of perturbation.\\ [0.3ex]
$\eta_l$  & client model learning rate. \\ [0.3ex]
$\eta_g$ & global model learning rate. \\[0.3ex]
$\tilde{\eta}$ & effective step-size. \\[0.3ex]
\multirow{2}{*}{$\theta^t$} & global server model at the $t$-th \\ & communication round. \\ [0.3ex]
$\theta_{i,k}^t$ & the $i$-th client model after $k$ mini-batch steps. \\ [0.3ex]
$g_\delta$ & gradients of weight with perturbation. \\ [0.3ex]
\multirow{2}{*}{$\epsilon$} &  a non-negative constant to bound gradients \\ & difference. \\ [0.3ex]
$\Gamma$ & overall non-iid drift term. \\ [0.3ex]
\hline
\end{tabular}
\caption{Notations occurred in the paper.}
\label{app:table:notation}
\end{table}

\section{Theoretical analysis of bounded drift}
\label{app:proof}

We give the theoretical analysis and proofs of HarmoFL in this section. First, we state the convexity, smoothness, and bounder gradient assumptions about the local function $F_i$ and global function $F$, which are typically used in optimization
literature~\cite{fedprox,fedadam,pmlr-v119-karimireddy20a,tong2020effective}.
\subsection{Assumptions}

\begin{assumption}($\beta\text{-smooth}$)
\label{assumption:1}
\{$F_i$\} are $\beta\text{-smooth}$ and satisfy
$$
\left\|\nabla F_{i}(\theta)-\nabla F_{i}(\theta_i)\right\| \leq \beta\|\theta-\theta_i\|, \text { for any } i, \theta, \theta_i
$$
\end{assumption}

\begin{assumption}(bounded gradient dissimilarity)
\label{assumption:2}
\text{there exist constants $G \geq 0$ and $B \geq 1$ such that}
$$
\frac{1}{N} \sum_{i=1}^{N}\left\|\nabla F_{i}(\theta)\right\|^{2} \leq G^{2}+B^{2}\|\nabla F(\theta)\|^{2}, \forall \theta
$$
\text{if \{$F_i$\} are convex, we can relax the assumption to}
$$
\frac{1}{N} \sum_{i=1}^{N}\left\|\nabla F_{i}(\theta)\right\|^{2} \leq G^{2}+ 2\beta B^{2}(F(\theta) - F^*), \forall \theta
$$
\end{assumption}

\begin{assumption}(bounded variance)
\label{assumption:3}
$g_{i}(\theta):=\nabla F_{i}(\theta ; x_i)$ 
\text{is unbiased stochastic gradient of $F_i$ with bounded variance}

$$
\E_{x_i} [\left\|g_{i}(\theta)-\nabla F_{i}(\theta)\right\|^{2}] \leq \sigma^{2}, \text{ for any } i, \theta
$$
\end{assumption}

\begin{lemma}(relaxed triangle inequality)
\label{lemma:tri_ineq}
Let ${a_1, \cdots , a_n }$ be $n$ vectors in $\R^d$. Then the following are true:
\begin{align*}
1. \left\|\boldsymbol{a}_{i}+\boldsymbol{a}_{j}\right\|^{2} \leq(1+\gamma)\left\|\boldsymbol{a}_{i}\right\|^{2}+\left(1+\frac{1}{\gamma}\right)\left\|\boldsymbol{a}_{j}\right\|^{2} \\
\text { , for any } \gamma > 0
\end{align*}

\begin{align*}
2. \|\sum_{i=1}^{n} \boldsymbol{a}_{i}\|^{2} \leq n \sum_{i=1}^{n}\left\|\boldsymbol{a}_{i}\right\|^{2}.
\end{align*}
\end{lemma}
\textit{Proof:} The proof of the first statement for any $\gamma > 0$ follows from the identity:
\begin{align*}
&\left\|\boldsymbol{a}_{i}+\boldsymbol{a}_{j}\right\|^{2}=\\
&\quad(1+\gamma)\left\|\boldsymbol{a}_{i}\right\|^{2}+\left(1+\frac{1}{\gamma}\right)\left\|\boldsymbol{a}_{j}\right\|^{2}-\left\|\sqrt{\gamma} \boldsymbol{a}_{i}+\frac{1}{\sqrt{\gamma}} \boldsymbol{a}_{j}\right\|^{2}.
\end{align*}
For the second inequality, we use the convexity of $x \rightarrow \|x\|^2$ and Jensen’s inequality
$$
\left\|\frac{1}{n} \sum_{i=1}^{n} \boldsymbol{a}_{i}\right\|^{2} \leq \frac{1}{n} \sum_{i=1}^{n}\left\|\boldsymbol{a}_{i}\right\|^{2}.
$$
\begin{lemma}(separating mean and variance)
\label{lemma:sep_var}
Let $\left\{\Xi_{1}, \ldots, \Xi_{\tau}\right\}$ be $n$ random variables in $\R^d$ which are not necessarily independent. First suppose that their mean is $\E\left[\Xi_{i}\right]=\xi_{i}$ and variance is bounded as $\E\left[\Xi_{i} - \xi_{i} \right] \leq \sigma^2$. Then, the following holds
$$
\E [\|\sum_{i=1}^{n} \Xi_{i}\|^{2}] \leq \|\sum_{i=1}^{n} \xi_{i}\|^{2} + n^{2} \sigma^{2}.
$$
\end{lemma}
\textit{Proof.} For any random variable $X$, $\E[X^2] = (\E[X-\E[X]])^2 + (\E[X])^2$ implying 
$$
\E [\|\sum_{i=1}^{n} \Xi_{i}\|^{2}] = \|\sum_{i=1}^{n} \xi_{i}\|^{2} + \E[\| \sum_{i=1}^{n} \Xi_i - \xi_i \|^2],
$$
expand the above equation using Lemma~\ref{lemma:tri_ineq}, we can have
$$
\E[\| \sum_{i=1}^{n} \Xi_i - \xi_i \|^2] \leq  \tau \sum_{i=1}^n \E[\|\Xi_i - \xi_i\|^2] \leq n^2\sigma^2.
$$

\subsection{Theorem of bounded drift and proof}
We first restate the Theorem~\ref{thm:bounded_drift} with some additional details and then give the proof.
Recall that $\Gamma=\frac{1}{KN}\sum_{k=1}^K \sum_{i=1}^N \E [ \|\theta_{i,k}^t - \theta^t \|^2 ]$.
\begin{theorem}
\label{app:thm:bounded_drift}
Suppose that the functions $\{F_i\}$ satisfies assumptions~\ref{assumption:1},~\ref{assumption:2} and ~\ref{assumption:3}. Denote the effective step-size $\tilde{\eta}=K\eta_g\eta_l$, then the updates of HarmoFL have a bounded drift:

\begin{itemize}
\item Convex:
\begin{align*}
\Gamma &\leq \frac{4\tilde{\eta}^2G^2}{\eta_g^2} + \frac{4\tilde{\eta}^2\epsilon^2(N-1)^2}{\eta_g^2 N^2} + \frac{2\tilde{\eta}^2\sigma^2}{K\eta_g^2} \\
&\quad\quad + \frac{8\beta\tilde{\eta}^2B^2}{\eta_g^2}(F(\theta) - F^*).
\end{align*}
\item Non-convex:
\begin{align*}
\Gamma &\leq
\frac{4\tilde{\eta}^2(G^2+B^2\|\nabla F(\theta)\|^2)}{\eta_g^2} + \frac{4\tilde{\eta}^2\epsilon^2(N-1)^2}{\eta_g^2 N^2} \\ 
&\quad\quad + \frac{2\tilde{\eta}^2\sigma^2}{K\eta_g^2}.
\end{align*}
\end{itemize}
\end{theorem}

\paragraph{Proof:}
First, we consider $K=1$, then this theorem trivially holds, since $\theta_{i,0} = \theta$ for all $i \in [N]$. Then we assume $K \ge 2$, and have 
\begin{align*}
\E\|\theta_{i,k} - \theta\|^2 &= \E\|\theta_{i,k-1} - \theta - \eta_l g_i(\theta_{i,k-1})\|^2 \\
&\leq \E\|\theta_{i,k-1} - \theta - \eta_l \nabla F_i(\theta_{i,k-1})\|^2 + \eta_l^2\sigma^2 \\
&\leq (1+\frac{1}{K-1})\E\|\theta_{i,k-1} - \theta \|^2  \\ 
& \quad\quad + K\eta_l^2\|\nabla F_i(\theta_{i,k-1})\|^2 + \eta_l^2\sigma^2\\
&= (1+\frac{1}{K-1})\E\|\theta_{i,k-1} - \theta \|^2  \\ 
& \quad\quad + \frac{\tilde{\eta}^2}{\eta_g^2 K}\|\nabla F_i(\theta_{i,k-1})\|^2 + \frac{\tilde{\eta}^2\sigma^2}{K^2\eta_g^2} \\
&\leq (1+\frac{1}{K-1})\E\|\theta_{i,k-1} - \theta \|^2  \\ 
& \quad\quad + \frac{2\tilde{\eta}^2}{\eta_g^2 K}\|\nabla F_i(\theta_{i,k-1}) - \nabla F_i(\theta)\|^2 \\
&\quad\quad + \frac{2\tilde{\eta}^2}{\eta_g^2 K}\|\nabla F_i(\theta)\|^2  +\frac{\tilde{\eta}^2\sigma^2}{K^2\eta_g^2} \\
&\leq (1+\frac{1}{K-1})\E\|\theta_{i,k-1} - \theta \|^2  \\ 
& \quad\quad + \frac{2\tilde{\eta}^2}{\eta_g^2 K}(\frac{N-1}{N}\epsilon)^2 \\
&\quad\quad + \frac{2\tilde{\eta}^2}{\eta_g^2 K}\|\nabla F_i(\theta)\|^2
+\frac{\tilde{\eta}^2\sigma^2}{K^2\eta_g^2}.
\end{align*}
For the first inequality, we separate the mean and variance, the second inequality is obtained by using the relaxed triangle inequality with $\gamma=\frac{1}{K-1}$. For the next equality, it is obtained with the definition of $\tilde{\eta}$ and the rest inequalities follow the assumptions on gradients. Up to now, we get the recursion format and then we unroll this and get below:
\begin{align*}
\E\|\theta_{i,k} - \theta\|^2 &\leq \sum_{\omega=1}^{k-1}(1+\frac{1}{K-1})^\omega(\frac{2\tilde{\eta}^2}{\eta_g^2 K}\|\nabla F_i(\theta)\|^2  \\
& \quad\quad + \frac{\tilde{\eta}^2\sigma^2}{K^2\eta_g^2} + \frac{2\tilde{\eta}^2\epsilon^2(N-1)^2}{\eta_g^2 KN^2})\\
&\leq 2K(\frac{2\tilde{\eta}^2}{\eta_g^2 K}\|\nabla F_i(\theta)\|^2 + \frac{2\tilde{\eta}^2\epsilon^2(N-1)^2}{\eta_g^2 KN^2}) \\
& \quad\quad  + \frac{\tilde{\eta}^2\sigma^2}{K^2\eta_g^2},
\end{align*}
averaging over $i$ and $k$, if $\{F_i\}$ are non-convex, we can have
\begin{align*}
\Gamma & \leq \frac{1}{N}\sum_{i=1}^N \frac{4\tilde{\eta}^2}{\eta_g^2}\|\nabla F_i(\theta)\|^2 + \frac{4\tilde{\eta}^2\epsilon^2(N-1)^2}{\eta_g^2 N^2} + \frac{2\tilde{\eta}^2\sigma^2}{K\eta_g^2}\\
&\quad\quad \leq \frac{4\tilde{\eta}^2G^2}{\eta_g^2} + \frac{4\tilde{\eta}^2\epsilon^2(N-1)^2}{\eta_g^2 N^2} + \frac{2\tilde{\eta}^2\sigma^2}{K\eta_g^2} \\
&\quad\quad + \frac{4\tilde{\eta}^2B^2}{\eta_g^2}\|\nabla F(\theta)\|^2.
\end{align*}
if $\{F_i\}$ are convex, then we have 
\begin{align*}
\Gamma &\leq \frac{4\tilde{\eta}^2G^2}{\eta_g^2} + \frac{4\tilde{\eta}^2\epsilon^2(N-1)^2}{\eta_g^2 N^2} + \frac{2\tilde{\eta}^2\sigma^2}{K\eta_g^2} \\
&\quad\quad + \frac{8\beta\tilde{\eta}^2B^2}{\eta_g^2}(F(\theta) - F^*)
\end{align*}

\section{Complete experiment details and results}
\label{app:complete_exp}
In this section, we demonstrate details of the experimental setting and more results. For all experiments, we implement the framework with PyTorch library of Python version 3.6.10, and train and test our models on TITAN RTX GPU. The randomness of all experiments is controlled by setting three random seeds, i.e., 0, 1, and 2. 
\subsection{Experimental details}
\textbf{Breast cancer histology image classification.}
For the breast cancer histology image classification experiment, we use the all data from the public dataset Camelyon17~\cite{bandi2018detection} dataset. This dataset comprises 450,000 patches of breast cancer metastasis in lymph node sections from 5 hospitals. The task is to predict whether a given region of tissue contains any tumor tissue or not. All the data are pre-processed into the shape of $96\times96\times3$, for all clients, we use 20\% of client data for test, for the reaming data, we use 80\% for train and 20\% for validation. For training details, we use the DenseNet121~\cite{huang2017densely} and train it with a learning rate of 0.001 for 100 epochs, the batch size is 128. We use cross-entropy loss and SGD optimizer with the momentum of $0.9$ and weight decay of $1e^{-4}$.\\ 
\textbf{Histology nuclei segmentation.}
For the histology nuclei segmentation, we use three public datasets, including MoNuSAC2020~\cite{verma2021monusac2020}, MoNuSAC2018~\cite{kumar2019multi} and TNBC~\cite{naylor2018segmentation}, and we further divide the MoNuSAC2020 into 4 clients. The criterion is according to the official multi-organ split, where each organ group contains several specific hospitals and has no overlap with other groups. All images are reshaped to $256\times256\times3$. For all clients, we follow the original train-test split and split out 20\% of training data for validation. For the training process, we use U-Net~\cite{ronneberger2015u} and train the model for 500 epochs using segmentation Dice loss. We use the SGD optimizer with the momentum of $0.9$ and weight decay of $1e^{-4}$. The batch size is $8$ and the learning rate is $1e^{-4}$.\\
\textbf{Prostate MRI segmentation.}
For the prostate MRI segmentation, we use a multi-site prostate segmentation dataset~\cite{liu2020ms} containing 6 different data sources~\cite{nciisbi2013,lemaitre2015computer,litjens2014evaluation}. For all clients, we use 20\% of client data for test, for the reaming data, we use 80\% for train and 20\% for validation. We use cross-entropy and dice loss together to train the U-Net and use Adam optimizer with a beta of $(0.9,0.99)$ and weight decay of $1e^{-4}$. The communication rounds are the same as the nuclei segmentation task. Following \cite{liu2020ms}, we randomly rotate the image with the angle of $\{0,90,180,270\}$ and horizontally or vertically flip images.

\subsection{Ablation studies}
In this section we present the ablation experiments regarding our proposed two parts on all three datasets (see results in Tables~\ref{tb:ablation_class},~\ref{tb:ablation_seg1},~\ref{tb:ablation_seg2}). We start with the baseline method FedAvg and then add the amplitude normalization (AmpNorm), on top of this, we further utilize the weight perturbation (WeightPert) and thus completing our proposed whole framework HarmoFL. From all three tables can be observed that the amplitude normalization (the second row) exceeds the baseline with consistent performance gain on all clients for all three datasets, demonstrating the benefits of harmonizing local drifts. On top of this, adding the weight perturbation (the third row) to mitigate the global drifts can further boost the performance with a clear margin. 
\begin{table}[!h]
\small
\centering
\setlength\tabcolsep{3pt}
\caption{Ablation studies for the classification task.}
\vspace{-2mm}
\begin{tabular}{@{}ccccccccc@{}}
\toprule
\multirow{3}{*}{\begin{tabular}[c]{@{}c@{}}Baseline\\FedAvg\end{tabular}} & \multirow{3}{*}{\begin{tabular}[c]{@{}c@{}}Amp\\Norm\\\scriptsize{(Local)}\end{tabular}} & \multirow{3}{*}{\begin{tabular}[c]{@{}c@{}}Weight\\ Pert\\\scriptsize{(Global)}\end{tabular}} & \multicolumn{6}{c}{\begin{tabular}[c]{@{}c@{}}Breast Cancer Histology Image \\ Classification (Accuracy \%)\end{tabular}} \\ \cmidrule(l){4-9} 
& & & A & B & C & D & E & Avg \\ \midrule
\ding{51} & & & 91.6 & 84.6 & 72.6 & 88.3 & 78.4 & 83.1\\
\ding{51} & \ding{51} & & 94.3 & 90.6 & 90.4 & 95.1 & 93.0 & 92.6 \\
\ding{51} & \ding{51} & \ding{51} & 96.8 & 94.2 & 95.9 & 95.8 & 96.9 & 95.9 \\ \bottomrule
\end{tabular}
\label{tb:ablation_class}
\end{table}
\vspace{-2mm}
\begin{table}[!h]
\small
\centering
\setlength\tabcolsep{2.5pt}
\caption{Ablation studies for the nuclei segmentation task.}
\vspace{-2mm}
\begin{tabular}{@{}cccccccccc@{}}
\toprule
\multirow{3}{*}{\begin{tabular}[c]{@{}c@{}}Baseline\\FedAvg\end{tabular}} & \multirow{3}{*}{\begin{tabular}[c]{@{}c@{}}Amp\\Norm\\\scriptsize{(Local)}\end{tabular}} & \multirow{3}{*}{\begin{tabular}[c]{@{}c@{}}Weight\\Pert\\\scriptsize{(Global)}\end{tabular}} & \multicolumn{7}{c}{\begin{tabular}[c]{@{}c@{}} Nuclei Segmentation (Dice \%)\end{tabular}}\\ \cmidrule(l){4-10} 
& & & A & B & C & D & E & F & Avg\\[1ex] \midrule
\ding{51} & & & 73.4&72.9&72.5&69.0&67.6&50.0&67.6  \\
\ding{51} & \ding{51} & & 74.3&74.8&75.2	&74.9&	72.3&	64.6	&72.7\\
\ding{51} & \ding{51} & \ding{51} & 75.4&	75.8	&76.8&	77.0	&74.1&	68.1&	74.5 \\
\bottomrule
\end{tabular}
\label{tb:ablation_seg1}
\end{table}
\vspace{-3.5mm}
\begin{table}[!h]
\small
\centering
\setlength\tabcolsep{2.5pt}
\caption{Ablation studies for the prostate segmentation task.}
\vspace{-2mm}
\begin{tabular}{@{}cccccccccc@{}}
\toprule
\multirow{3}{*}{\begin{tabular}[c]{@{}c@{}}{Baseline}\\{FedAvg}\end{tabular}} & \multirow{3}{*}{\begin{tabular}[c]{@{}c@{}}{Amp}\\{Norm}\\\scriptsize{(Local)}\end{tabular}} & \multirow{3}{*}{\begin{tabular}[c]{@{}c@{}}{Weight}\\Pert\\\scriptsize{(Global)}\end{tabular}} & \multicolumn{7}{c}{\begin{tabular}[c]{@{}c@{}} Prostate MRI Segmentation (Dice \%)\end{tabular}}\\ \cmidrule(l){4-10} 
& & & A & B & C & D & E & F & Avg \\[1ex] \midrule
\ding{51} & & & 
89.4	&94.1 & 92.3 & 91.4 & 90.3 & 91.4 & 91.5\\
\ding{51} & \ding{51} & & 
91.4 & 94.8 & 94.2 & 92.6 & 93.4 & 93.0 & 93.2\\
\ding{51} & \ding{51} & \ding{51} &
94.3 & 95.3 & 95.5 & 94.2 & 94.4 & 94.2 & 94.7
\\ \bottomrule
\end{tabular}
\label{tb:ablation_seg2}
\end{table}

\subsection{Additional results}

\begin{figure*}[t]
\centering
\includegraphics[width=0.93\textwidth]{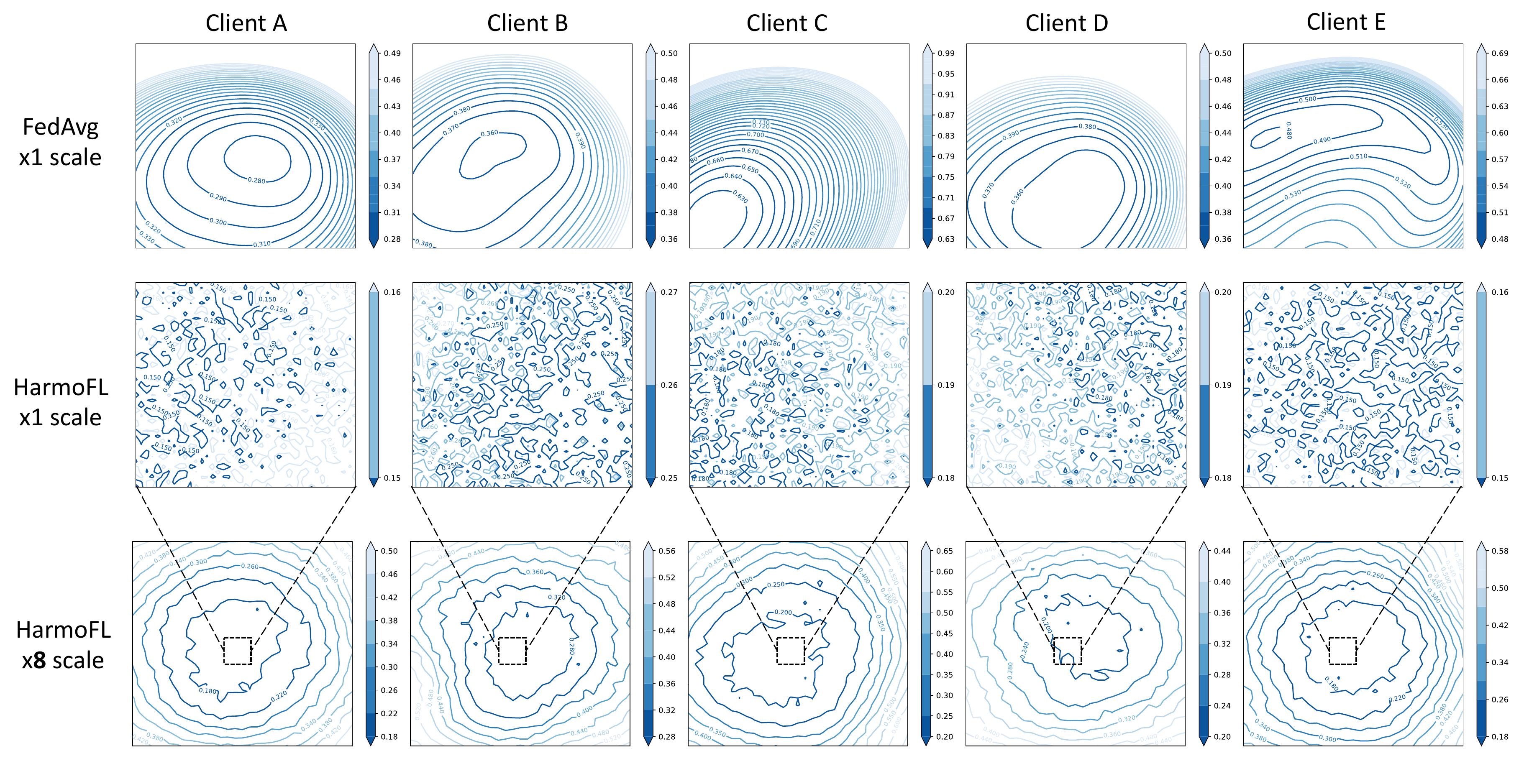}
\vspace{-5mm}
\caption{Loss landscape visualization of 5 clients in the breast cancer histology image classification task with different parameter changing scale. Our method is more robust to model parameter change, achieving a centered and flat area for both global solution and 5 local solutions, even with 8 times larger changing scales.}
\label{app:full_loss_2d}
\end{figure*}

\begin{figure}[!h]
\centering
\includegraphics[width=0.95\columnwidth]{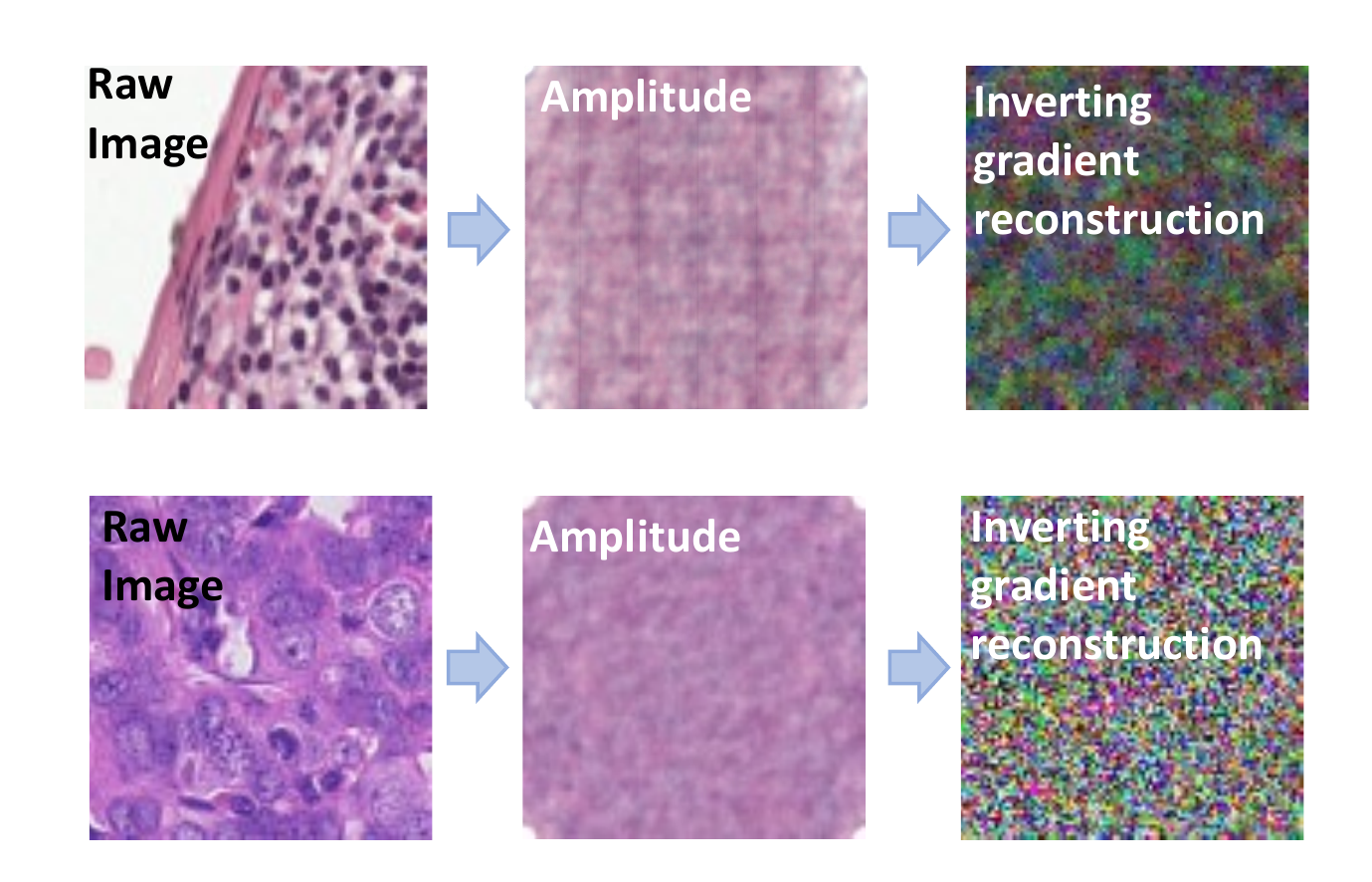} 
\vspace{-3mm}
\caption{Examples of inverting gradient attack using the amplitude extracted from original images.}
\label{app:inverse_grad}
\end{figure}

\begin{figure}[!h]
\centering
\includegraphics[width=0.95\columnwidth]{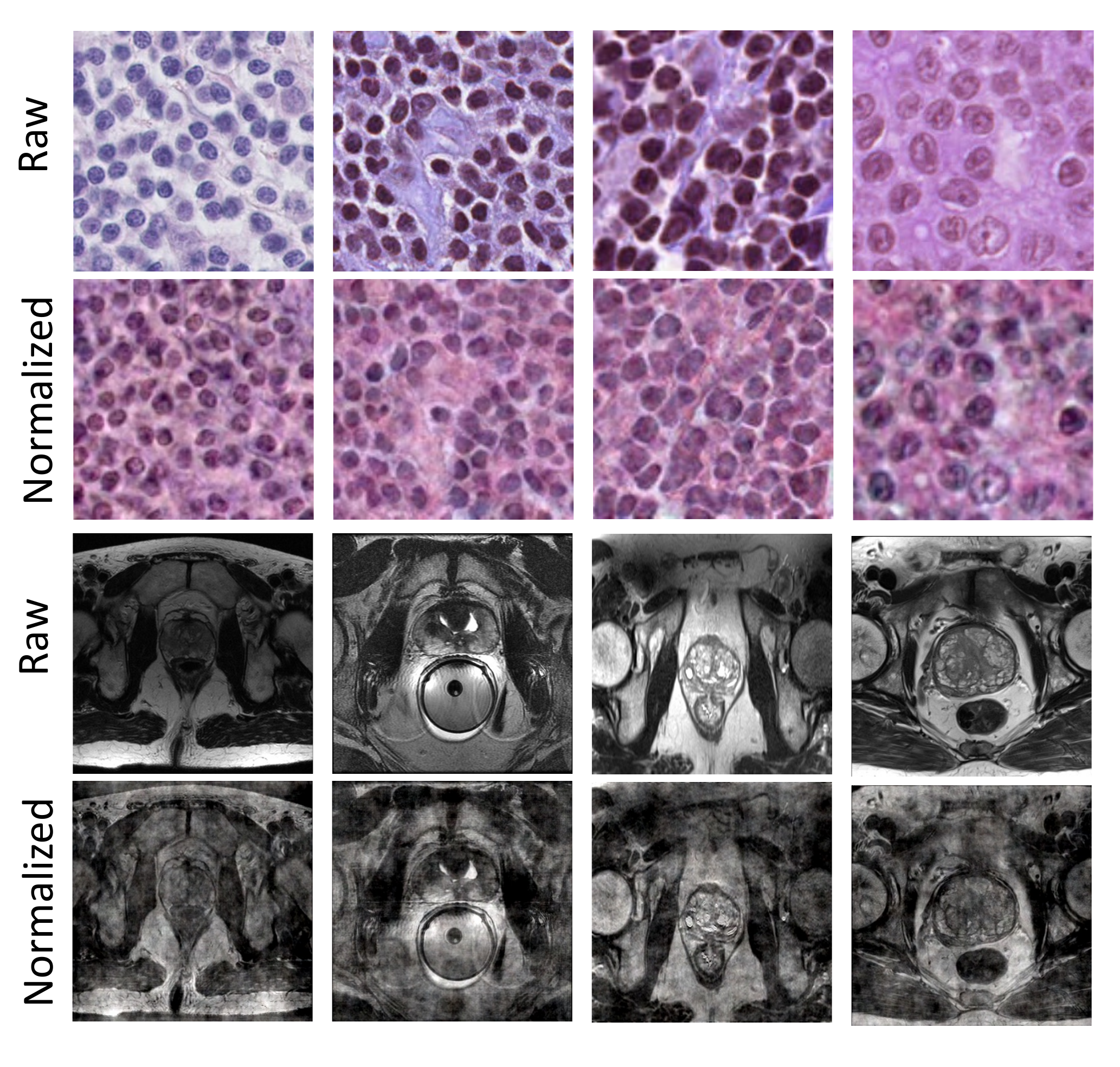} 
\vspace{-3mm}
\caption{Qualitative visualization results of using amplitude normalization.}
\label{app:amp_norm_vis}
\end{figure}

\begin{figure*}[t]
\centering
\includegraphics[width=0.99\textwidth]{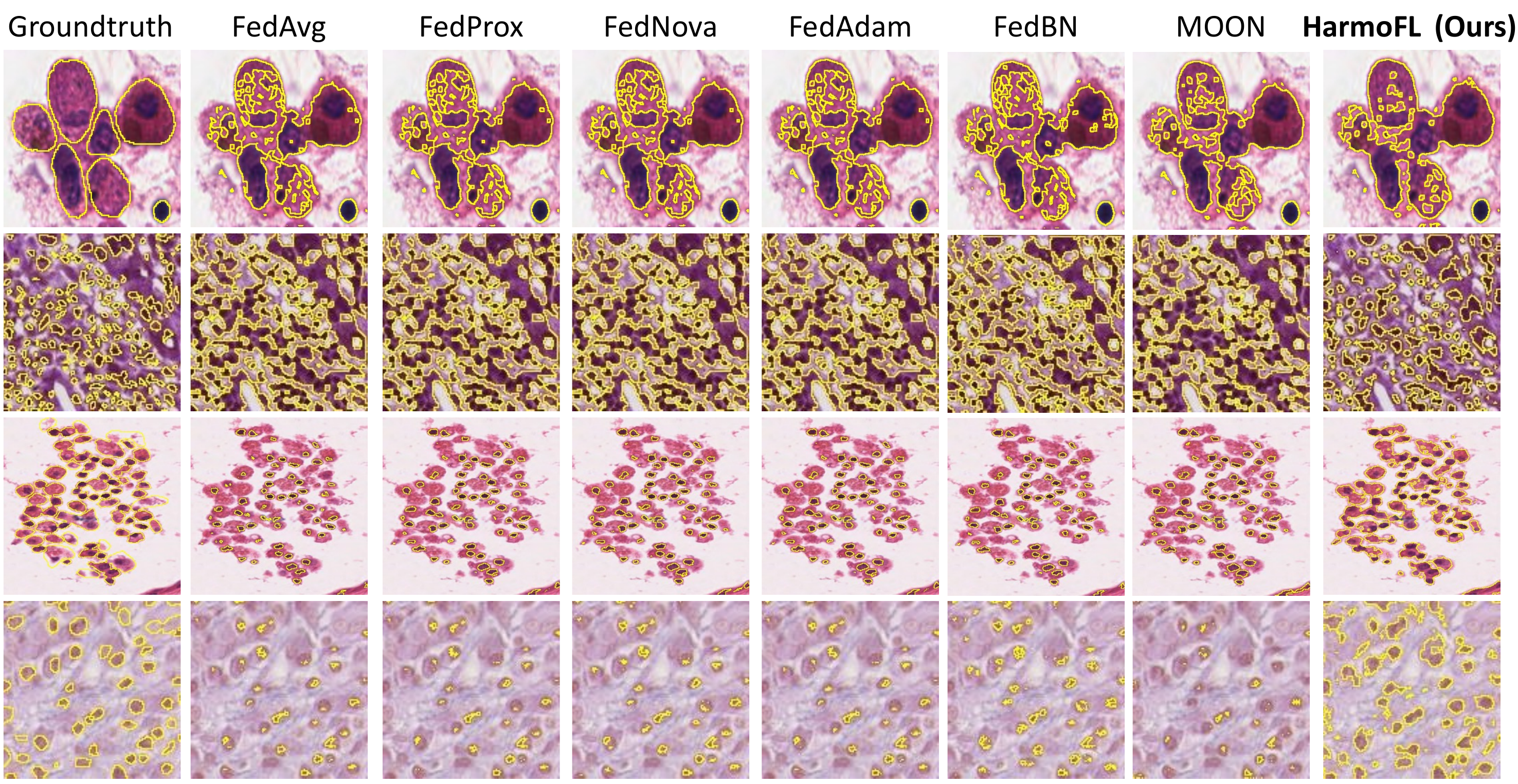}
\caption{Qualitative comparison on histology nuclei segmentation results with our method and other state-of-the-art methods.}
\label{app:add_nuclei_vis}
\end{figure*}

\begin{figure*}[t]
\centering
\includegraphics[width=0.99\textwidth]{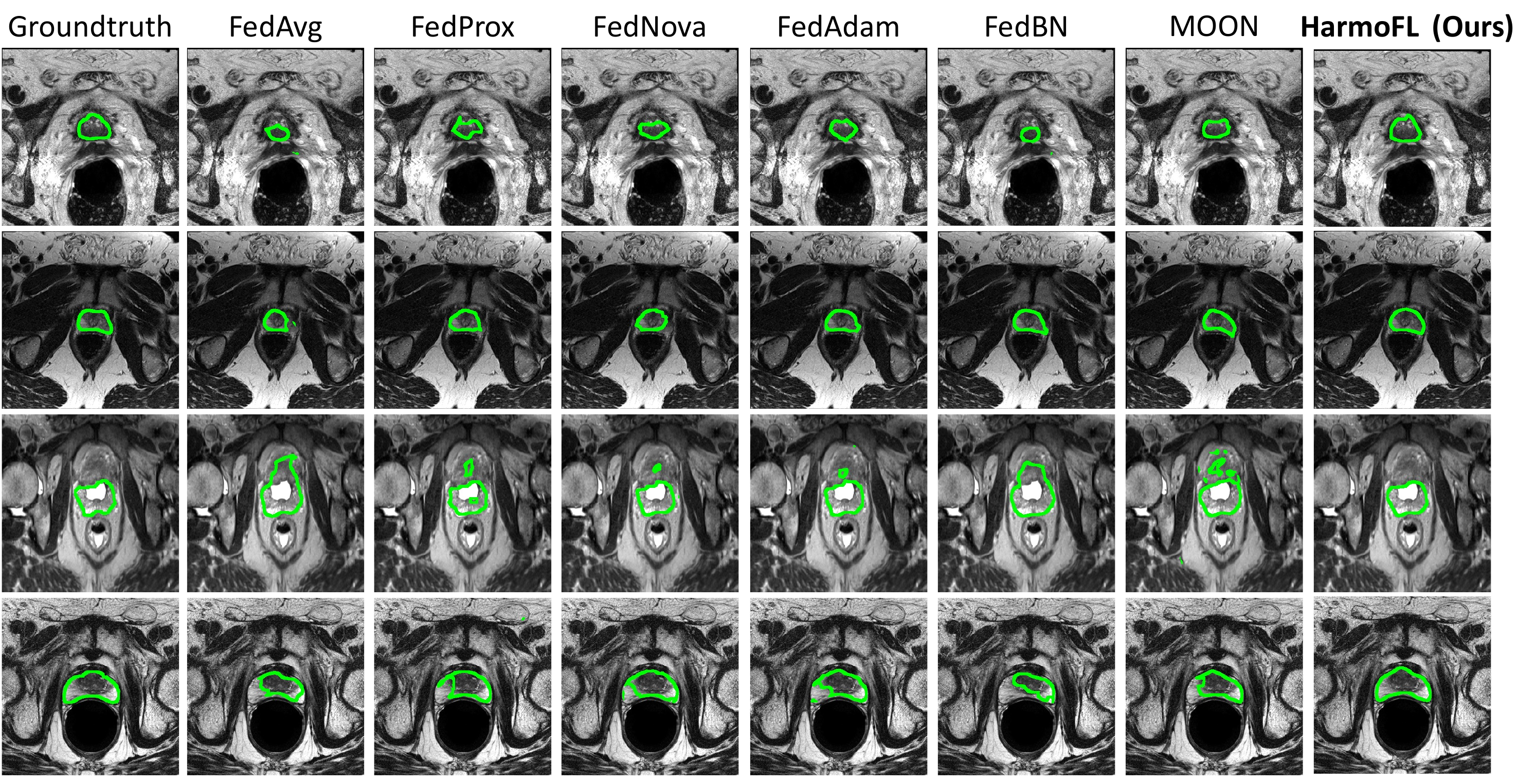}
\caption{Qualitative comparison on prostate MRI segmentation results with our method and other state-of-the-art methods.}
\label{app:add_prostate_vis}
\end{figure*}

\paragraph{Loss landscape visualization on breast cancer histology image classification.} We use the loss landscape visualization~\cite{visualloss} to empirically show the non-iid drift in terms of the $\Gamma_i := \frac{1}{K}\sum_{k=1}^K \E[\|\theta_{i,k}^t - \theta^t\|^2]$ for each client $i$. As shown in Fig.~\ref{app:full_loss_2d}, we draw the loss of FedAvg and our HarmoFL, each column represents a client. The 2-d coordination denotes a parameter space centered at the final obtained global model, loss value varies with the parameter change. From the first row can be observed that the model trained with FedAvg is more sensitive to a parameter change, and the minimal solution for each client does not match the global model (i.e., the center of the figure). The observation is consistent with the Introduction part, with the same objective function and parameter initialization for each client, the five local optimal solutions (local minima) have significantly different parameters, which is caused by the drift across client updates. Besides, the loss landscape shape of each client differs from others, making the global optimization fail to find a converged solution matching various local optimal solutions. However, from the second row of our proposed HarmoFL, we can observe the loss variation is smaller than FedAvg and robust to the parameter changes, showing flat solutions both locally and globally. We further draw loss values with a higher scale as shown in the third row, which means we change the parameter with a large range. From the third row can be observed that HarmoFL holds flat optima around the central area, even with 8 times wider parameter change range. Besides, the global solution well matches every client's local optimal solution, demonstrating the effectiveness of reducing drifts locally and globally.\\
\textbf{Gradient inversion attacks with the shared amplitudes across clients.}
To further explore the potential privacy issue regarding the inverting gradient attack by using our shared average amplitudes across client, we implement the inverting gradient method~\cite{geiping2020inverting} and show results in below Fig.~\ref{app:inverse_grad}. We first extract the amplitude from the original images and then conduct the inverting gradient attack on the amplitudes which will be shared during the federated communication. The reconstruction results of the attack are shown at the last column. It implies that it is difficult (if possible) to perfectly reconstruct the original data solely using frequency information.\\
\textbf{Visualization of amplitude normalization results.}
In this section we demonstrate the visual effects of our amplitude normalization. The qualitative results for the two segmentation datasets (Fig.~\ref{fig:amp_norm} is the classification dataset) are shown in Fig.~\ref{app:amp_norm_vis}. Given that amplitudes mainly relate to low-level features, the AmpNorm operation would not affect the high-level semantic structures. In addition, though the normalization might introduce tiny changes in some ambiguous boundaries, the network with a large receptive field would have the ability to combat such local changes by using contextual information.\\
\textbf{Visualization of segmentation results.} More qualitative segmentation results comparison on both prostate MRI segmentation and histology nuclei segmentation tasks are shown in Fig.~\ref{app:add_nuclei_vis} and Fig.~\ref{app:add_prostate_vis}. From Fig.~\ref{app:add_nuclei_vis}, as can be observed from the first row, alternative methods show several small separate parts for the segmentation on big cells while HarmoFL gives more complete segmentation results. As for small cells, HarmoFL has the capability to split each small region out while others may connect small nuclei together or only cover a very small explicit region. Specifically for the fourth row, others fail to segment some nuclei which have fewer differences with background, but our method demonstrates more accurate results. For the prostate MRI segmentation results shown in Fig.~\ref{app:add_prostate_vis}, the images from different hospitals show a shift in feature distributions, which brings difficulties for other methods to obtain a more accurate boundary. Especially for the third row, the compared methods fail to figure out the structure, while ours delineates the accurate boundary.

\end{document}